\documentclass{aa}
\usepackage[utf8]{inputenc}
\usepackage[colorlinks]{hyperref}
\usepackage{amsmath}
\usepackage{amssymb}
\usepackage{microtype}
\usepackage[T1]{fontenc}
\usepackage{lmodern}
\usepackage{natbib}
\usepackage{graphicx}
\usepackage{microtype}
\usepackage{color}
\usepackage[dvipsnames]{xcolor}
\usepackage{multirow}
\usepackage{xspace}
\usepackage[normalem]{ulem}

\bibpunct{(}{)}{;}{a}{}{,} 

\def\nustar{\textit{NuSTAR}}
\def\srg{\textit{SRG}}
\def\art{ART-XC}
\def\erosita{eROSITA}
\def\int{\textit{INTEGRAL}}
\def\urd29{$\rm{URD_{29}}$}
\def\ms6{$\rm{MS_{6}}$}

\def\arcmin{\hbox{$^\prime$}}
\def\arcsec{\hbox{$^{\prime\prime}$}}
\def\flux{erg\,s$^{-1}$\,cm$^{-2}$}

\def\lum{erg s$^{-1}$}

\def\aap{A\&A}

\def\apjs{ApJS}

\def\apjl{ApJLett}

\def\mysim{\mathord{\sim}}
\def\deg{^{\circ}}

\def\arcmin{\hbox{$^\prime$}}
\def\arcsec{\hbox{$^{\prime\prime}$}}
\def\flux{erg\,s$^{-1}$\,cm$^{-2}$}

\def\lum{erg s$^{-1}$}

\def\Lx{$L_{\rm X}$}

\begin{document}
\title{The ART-XC telescope on board the \srg\ observatory}

\author{
  M.~Pavlinsky\inst{1} \and A.Tkachenko\inst{1} \and V. Levin\inst{1} \and N. Alexandrovich\inst{1} \and V. Arefiev\inst{1} \and V. Babyshkin\inst{2} \and O. Batanov\inst{1} \and Yu. Bodnar\inst{3} \and A. Bogomolov\inst{1} \and A. Bubnov\inst{1} \and M. Buntov\inst{1} \and R. Burenin\inst{1} \and I. Chelovekov\inst{1} \and C.-T. Chen\inst{4} \and T. Drozdova\inst{1} \and S. Ehlert\inst{5} \and E. Filippova\inst{1} \and S. Frolov\inst{3} \and D. Gamkov\inst{1} \and S. Garanin\inst{3} \and M. Garin\inst{3} \and A. Glushenko\inst{1} \and A. Gorelov\inst{3} \and S. Grebenev\inst{1} \and S. Grigorovich\inst{3} \and P. Gureev\inst{2} \and E. Gurova\inst{1} \and R. Ilkaev\inst{3} \and I. Katasonov\inst{1} \and A. Krivchenko\inst{1} \and R. Krivonos\inst{1} \and F. Korotkov\inst{1} \and M. Kudelin\inst{1} \and M. Kuznetsova\inst{1} \and V. Lazarchuk\inst{3} \and I. Lomakin\inst{2} \and I. Lapshov\inst{1} \and V. Lipilin\inst{1} \and A. Lutovinov\inst{1}\thanks{E-mail: aal@iki.rssi.ru} \and I. Mereminskiy\inst{1} \and S. Molkov\inst{1} \and V. Nazarov\inst{1} \and V. Oleinikov\inst{1} \and E. Pikalov\inst{3} \and B. D. Ramsey\inst{5} \and I. Roiz\inst{3} \and A. Rotin\inst{1} \and E. Sankin\inst{3} \and A. Ryadov\inst{3} \and S. Sazonov\inst{1} \and D. Sedov\inst{3} \and A. Semena\inst{1} \and N. Semena\inst{1} \and D. Serbinov\inst{1} \and A. Shirshakov\inst{2} \and A. Shtykovsky\inst{1} \and A. Shvetsov\inst{3} \and R. Sunyaev\inst{1,6} \and D. A. Swartz\inst{4} \and V. Tambov\inst{1} \and V. Voron\inst{7} \and A. Yaskovich\inst{1}
}

\institute{
     Space Research Institute, 84/32 Profsouznaya str.,
     Moscow 117997, Russian Federation
\and Lavochkin Association, 24 Leningradskaya str.,
     Khimki 141400, Moscow Region, Russian Federation
\and Russian Federal Nuclear Center – All-Russian Scientific Research Institute of Experimental Physics (RFNC-VNIIEF), 37 Mira Ave, Sarov 607188, Nizhny
     Novgorod region, Russian Federation
\and Universities Space Research Association, Huntsville,
     AL 35805, USA
\and NASA/Marshall Space Flight Center, Huntsville,
     AL 35812 USA
\and Max-Planck-Institut f\"ur Astrophysik, Karl-Schwarzschild-Stra{\ss}e, D-85741 Garching, Germany
\and State Space Corporation Roscosmos, 42 Schepkina str.,
     Moscow 107996, Russian Federation
}

\abstract{\art\ (Astronomical Roentgen Telescope -- X-ray Concentrator) is the hard X-ray instrument with grazing incidence imaging optics on board the {\it Spektr-Roentgen-Gamma (SRG)} observatory. The \srg\  observatory is the flagship astrophysical mission of the Russian Federal Space Program, which was successively launched into orbit around the second Lagrangian point (L2) of the Earth–Sun system with a Proton rocket from the Baikonur cosmodrome on 13 July 2019. The \art\ telescope will provide the first ever true imaging all-sky survey performed with grazing incidence optics in the 4--30~keV energy band and will obtain the deepest and sharpest map of the sky in the energy range of 4--12~keV. Observations performed during the early calibration and performance verification phase as well as during the on-going all-sky survey that started on 12 Dec. 2019 have demonstrated that the in-flight characteristics of the \art\ telescope are very close to expectations based on the results of ground calibrations. Upon completion of its 4-year all-sky survey, \art\ is expected to detect $\sim 5000$ sources ($\sim 3000$ active galactic nuclei, including heavily obscured ones, several hundred clusters of galaxies, $\sim 1000$ cataclysmic variables and other Galactic sources), and to provide a high-quality map of the Galactic background emission in the 4--12~keV energy band. \art\ is also well suited for discovering transient X-ray sources. In this paper, we describe the telescope, results of its ground calibrations, major aspects of the mission, the in-flight performance of \art\ and first scientific results.
}
\keywords{Space vehicles: instruments – X-rays: general – Surveys}

\authorrunning{M.\,Pavlinsky et al.}

\maketitle

\section{Introduction}

The {\it Mikhail Pavlinsky}{\footnote{Mikhail Pavlinsky (1959-2020) had been PI for the \art\ telescope and Co-I for the \srg\ observatory. He made a decisive contribution to the successful realization of this project and the creation in Russia of a modern school of space instrumentation. To our greatest regret, in July 2020, M. Pavlinsky passed away. In his memory it was decided to name the \art\ telescope after M. Pavlinsky -- the {\it Mikhail Pavlinsky} \art\ telescope.}} Astronomical Roentgen Telescope -- X-ray Concentrator (\art, \citealt{2011SPIE.8147E..06P,2016SPIE.9905E..1JP}) is one of two X-ray telescopes of the \textit{Spektr-Roentgen-Gamma} (\srg) observatory -- the flagship astrophysical project of the Russian Federal Space Program (Sunyaev et al., 2020). \art\ is designed to produce true images of the X-ray sky using the grazing incidence imaging X-ray optics technique. \art\ can also be used in a mode with a much larger effective field of view but without good angular resolution (''concentrator mode''). The \art\ and \erosita\ \citep[the other instrument of the \srg\ observatory,][]{2020arXiv201003477P} telescopes complement each other, being sensitive in the 4--30~keV and 0.2--8~keV energy bands, respectively.

The only all-sky survey carried out previously with a grazing incidence X-ray telescope is the \textit{ROSAT} all-sky survey (RASS) in the 0.1--2.4~keV energy band \citep{1999A&A...349..389V}. At higher energies, all-sky X-ray surveys have been performed either with collimator instruments (e.g. \textit{UHURU}, \citealt{1978ApJS...38..357F}, and \textit{HEAO1},  \citealt{1984ApJS...56..507W}) or with coded-mask aperture telescopes (e.g. \textit{INTEGRAL}/IBIS and \textit{Swift}/BAT, see below). The sensitivity of these surveys was strongly limited by their poor angular resolution. The \art\ telescope will provide the first ever true imaging all-sky survey performed with grazing incidence optics in the 4--30~keV energy band, i.e. at significantly higher energies compared to RASS. \art\ is optimized for conducting surveys in the 4--12~keV band with maximum sensitivity at around 8--10~keV.

\art\ was developed by the Space Research Institute (IKI, Moscow) and the Russian Federal Nuclear Center -- All-Russian Scientific Research Institute for Experimental Physics (RFNC-VNIIEF, Sarov). The development of the X-ray optics for \art\ proceeded independently at VNIIEF and NASA's Marshall Space Flight Center (MSFC). The VNIIEF mirror systems were installed in the qualification model of \art, which was subjected to vibration and endurance tests. This led to a significant reduction of the \art\ project's execution time. The flight models of \art\ X-ray mirror systems were simultaneously developed, fabricated and calibrated by MSFC. It is important to note that the great experience, gained by IKI in the development and operating of the ART-P telescope on board the {\it GRANAT} observatory \citep[see, e.g.,][]{1993ApJ...407..606S,1994ApJ...425..110P} as well as in development and working with instruments of the {\it Roentgen} observatory on board the {\it MIR} space station \citep[see, e.g.,][]{1991SvAL...17..409S}, played a significant role in the creation of \art\ and formulation of its scientific tasks.

\art\ is the first Wolter grazing incidence X-ray telescope developed and launched into space by Russia.

\section{The \art\ telescope}

\subsection{Structure}
\label{sect:structure}
\begin{figure*}
\centerline{
\includegraphics[width=0.98\textwidth,bb=100 550 570 790,clip]{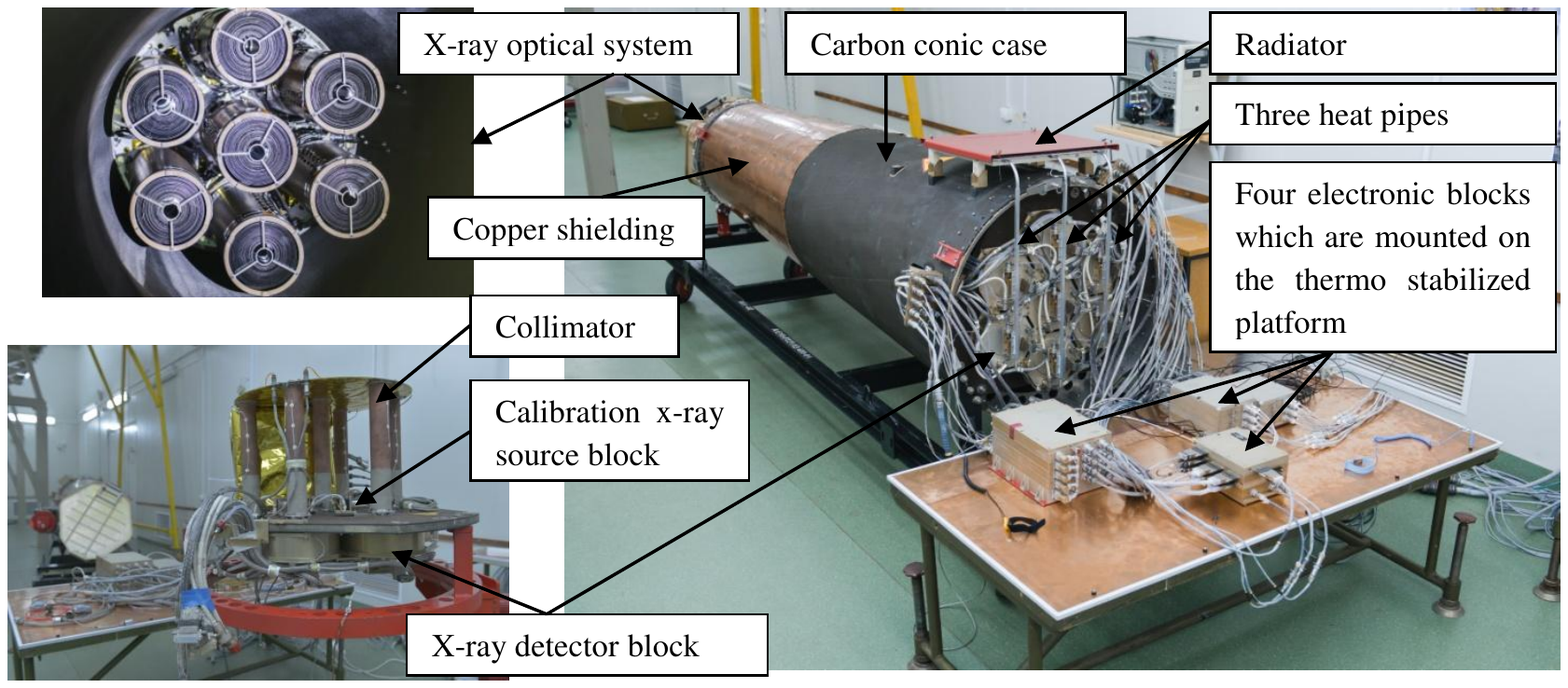}
}
\caption{The \art\ telescope. (a) Upper left: Face-on view of the 7 MSs looking towards the detectors. (b) Lower left: Side view of the detector block prior to assembly; there are 7 detectors located in the detector block that align with the 7 MSs. Above the detector block, in the direction toward the MSs, are collimators for each detector each with a calibration source block. (c) Right: Complete telescope housing all 7 MSs (far left in the image) and the detector block (near right) and associated hardware, including elements of the thermal control system -- radiator and heat pipes. The electronic blocks in foreground are to be mounted on a thermally-stabilized platform for flight (not shown). The star tracker is  located at the MS end of the carbon/copper telescope tube (hidden from view in this image).}
\label{fig:art_design}
\end{figure*}

\art\ consists of the telescope itself and four separate electronics units that are mounted on a thermo-stabilized platform located 0.5 meters under the telescope (Fig.~\ref{fig:art_design}, \citealt{2012SPIE.8443E..1TP}).

The telescope has a weight of about 350~kg, an approximate dimensions of 3.5~m of height and 0.9~m in diameter and power consumption of 150~W. The main elements of the telescope are seven X-ray mirror systems (MSs) and corresponding X-ray detectors. Each pair "mirror system -- detector" forms a "telescope module", these modules are co-aligned and being referred to below as T1...T7. The MSs are mounted on a platform situated in the upper part of the telescope structure. The seven X-ray detectors compose a focal plane assembly. The detectors and mirrors are enclosed in a carbon fiber conic housing. The upper part of the conic housing is covered with a copper shield. It blocks the side aperture stray light, which would otherwise significantly increase the overall detector X-ray background. This issue was investigated in detail by \cite{2017arXiv171102719M} for the \textit{NuSTAR} orbital telescope \citep{2013ApJ...770..103H}.

The telescope has an on-board calibration system, which is used to determine detector gain and energy resolution, and a thermal control system. The calibration system contains a drive control unit and seven calibration X-ray source blocks -- one for each detector. The calibration X-ray source consists of Am$^{241}$ and Fe$^{55}$ and is pulled out of the lead box using a stepper motor.

The telescope is a rather complex object from the point of view of ensuring its thermal stability. The telescope's thermal control system consists of 36 active elements -- heaters mounted in different places on the telescope's structure \citep{2014SPIE.9144E..4TS}. The instrument has two strictly thermally stabilized zones: X-ray mirrors and detectors. A stable temperature of mirror shells in the range of $20\pm2$\,$^{\circ}$C is provided by the heated outer mirror shell and the thermal mirror baffle. The lower part of the outer mirror shell has a temperature of 27\,$^{\circ}$C, while the upper part has 28\,$^{\circ}$C. The temperature of the thermal mirror baffle is 22\,$^{\circ}$C with the time stability of 0.01\,$^{\circ}$C. The characteristic temperatures of the detector units are discussed below.

The separate electronics units are the information collection and control unit, two blocks of electronics for detectors and the thermal control unit.

A BOKZ-MF star tracker is mounted on the MS platform next to the MSs.

Below we describe the systems and components of the \art\ telescope in detail. Key parameters of the instrument are summarized in Table\,\ref{tab:artxc_keypar}. The quoted values take into account the results of extensive ground calibrations, which are discussed in \S\ref{s:groundcal} below. As further discussed in \S\ref{s:inflight}, in-flight calibrations have generally confirmed the preflight parameters of \art.

\begin{table*}[]
\caption{\art\ characteristics.}\label{tab:artxc_keypar}
\centerline{\begin{tabular}{l|c}
\hline
\multicolumn{2}{c}{\art\ key parameters} \\
\hline
Telescope mass & 350~kg \\
Dimensions & $3.5\times\diameter0.9$~m \\
Power & 150~Watts \\
Energy range & 4--30~keV \\
Effective area for pointed observations & 
{385~cm$^2$} @ 8.1~keV$^*$ \\
Grasp 
Angular resolution (FWHM) in the survey mode & \\
(limited by detector pixel size) & $53\arcsec$ \\
Detector efficiency & $50$\%~@~4.6~keV, $86$\%~@~8.1~keV \\
Energy resolution & 9\% @ 13.9~keV$^{**}$ \\
Time resolution & 23 $\mu$s \\
Dead time & 0.77~ms \\
\hline
\hline
\multicolumn{2}{c}{\art\ Optics} \\
\hline
Number of MSs& 7 \\
Nominal focal length & 2700~mm \\
Defocusing & $-7$~mm \\
Number of nested mirror shells & 28 \\
Form of shell & Wolter-I \\
Diameters of shells (intersection) & 49--145~mm \\
Thickness of shells & 0.25--0.35~mm \\
Material of shells & Ni/Co \\
Mirror coating & Ir (90\% bulk density) \\
Entrance filter & 18.5~$\mu$m Mylar film with 0.11~$\mu$m Al layer\\
HPD on-axis, arcsec & 30--35$\arcsec^{***}$ \\
\hline
\hline
\multicolumn{2}{c}{\art\ Detectors} \\
\hline
Detector type & CdTe Schottky Diode double sided strip \\
 & (Acrorad, Japan) \\
CdTe crystal's size & $29.95\times29.95\times1.00$~mm \\
Working area & $28.56\times28.56$~mm  \\
Number of strips & $48\times48$ \\
Strip width & 520~$\mu$m \\
Inter-strip distance & 75~$\mu$m \\
ASIC, 2 pcs. & VA64TA1 (Ideas, Norway) \\
Working energy range & 4--120~keV  \\
Entrance window & Be, $\diameter30$~mm, thickness of 100~$\mu$m \\
\hline
\end{tabular}}
\flushleft
\hspace{15mm} $^*$ -- calibration Cu line\\
\hspace{15mm} $^{**}$ -- calibration $^{237}$Ne line\\
\hspace{15mm} $^{***}$ -- see Table \ref{tab:artxc_mirrpar}\\
\end{table*}

\subsection{Mirror systems}

\begin{figure}
\centerline{
\includegraphics[width=0.98\columnwidth,clip]{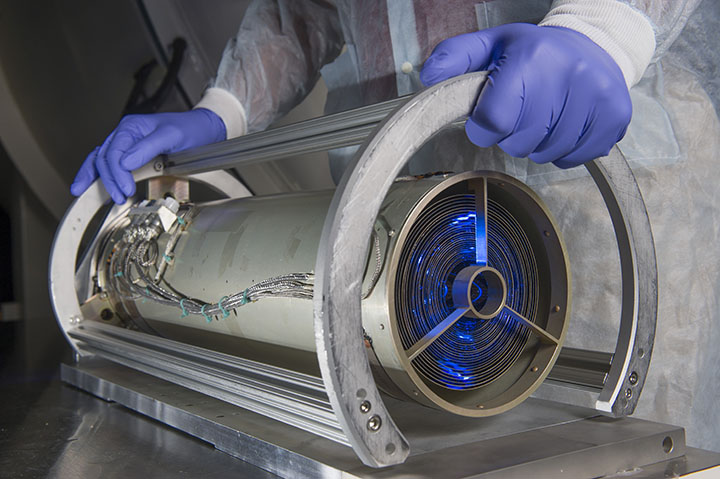}
}
\caption{One mirror system being prepared for testing at the MSFC Stray Light Test Facility.}
\label{fig:ms_msfc}
\end{figure}

The design of the \art\ X-ray optics was developed independently at VNIIEF and MSFC based on specifications by IKI. MSFC used the classical parabolic/hyperbolic shape of Wolter type I mirrors instead of a conical approximation to Wolter I geometry used by VNIIEF. The major characteristics, such as mass and effective area, of the MSs were similar, but the modules produced by MSFC had a significantly better angular resolution. VNIIEF fabricated seven MSs. All of them were installed in the qualification model of \art\, which was subjected to vibration and endurance tests. MSFC fabricated, tested and calibrated eight identical X-ray MSs for \art\ (see Fig.~\ref{fig:ms_msfc}) (\citealt{2012SPIE.8443E..1UG}, \citealt{2014SPIE.9144E..4UG}, \citealt{2017ExA....44..147K}). Seven of them were installed into the flight model of the telescope and the eighth became a spare.

\begin{figure}
\centerline{
\includegraphics[width=0.98\columnwidth,clip]{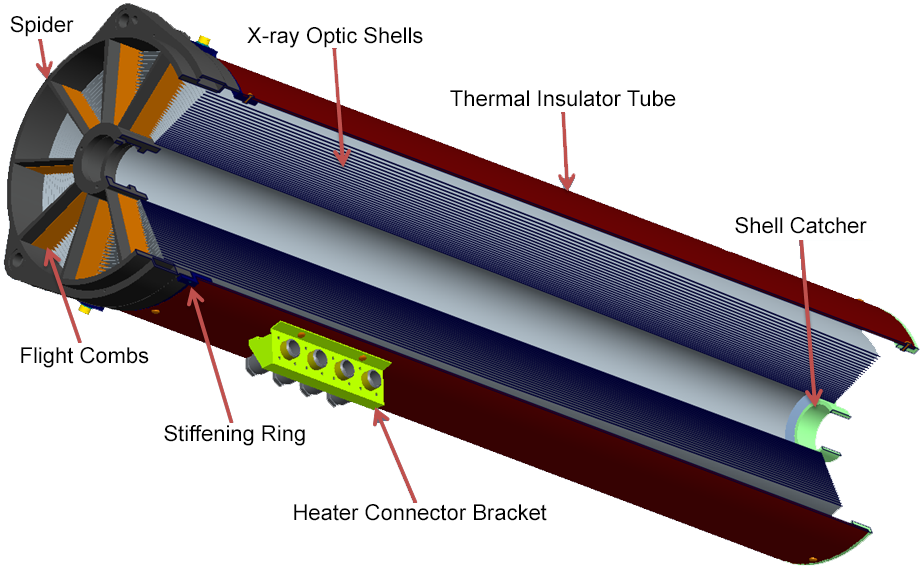}
}
\caption{Cross section of the mirror system. The inner baffle tube (for stray light reduction) and the heaters are not shown}
\label{fig:rzs_model}
\end{figure}

Each MS (see Fig.~\ref{fig:rzs_model}) contains 28 Wolter-I nested mirror shells. The shells were fabricated using an electroformed-nickel-replication technique. The shells have diameters ranging from 49 to 145~mm. Their thickness varies with radius from 250 to 350~$\mu$m. The outer shells are made thicker, which increases their rigidity and, hence, improves the angular resolution of the MS. The inner surface of the \art\ nickel-cobalt mirrors is coated with a $\mysim10$~nm layer of 90\% bulk density iridium (Ir). This metal has an X-ray reflective index higher than that of gold at energies above 10~keV. The upper ends of the shells are glued into a supporting ``spider''.

\begin{figure}
\centerline{
\includegraphics[width=0.98\columnwidth,clip]{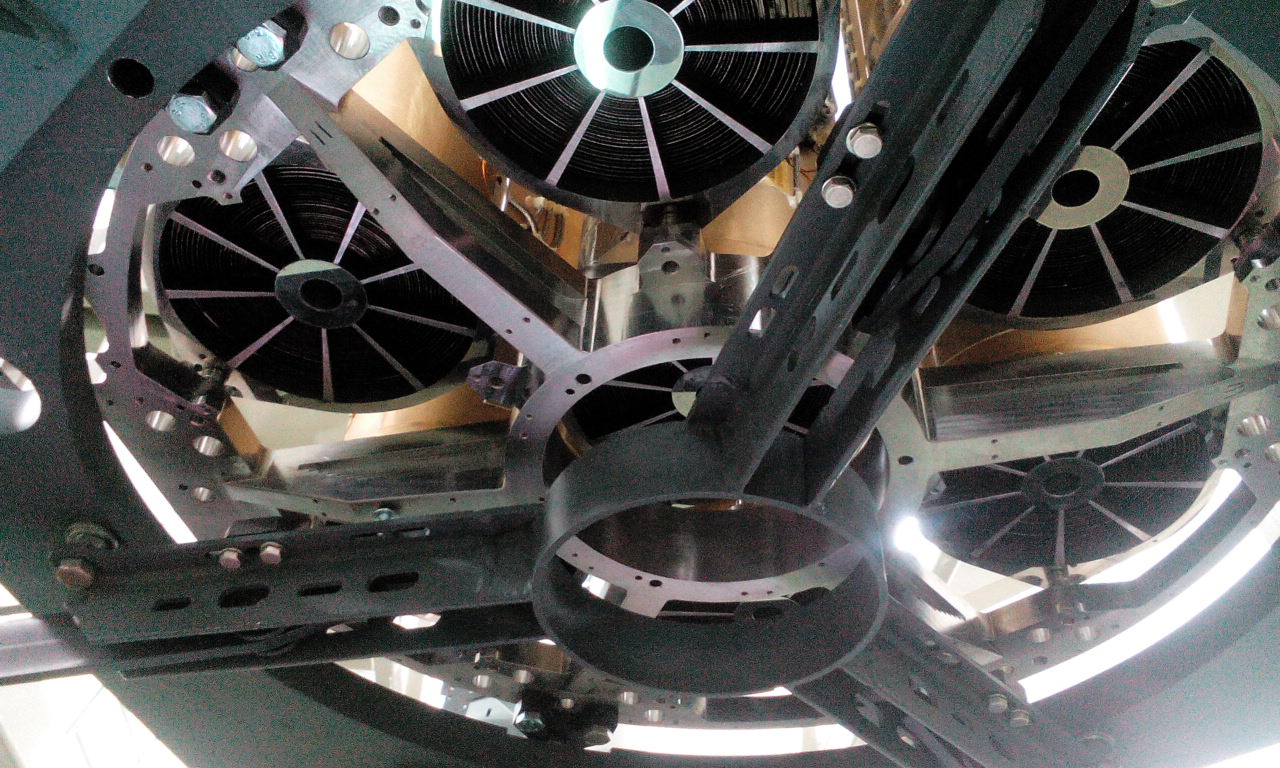}
}
\caption{Mirror systems installed at the assembly stand.}
\label{fig:ms_sarov}
\end{figure}

\begin{figure}
\centerline{
\includegraphics[width=0.98\columnwidth,clip]{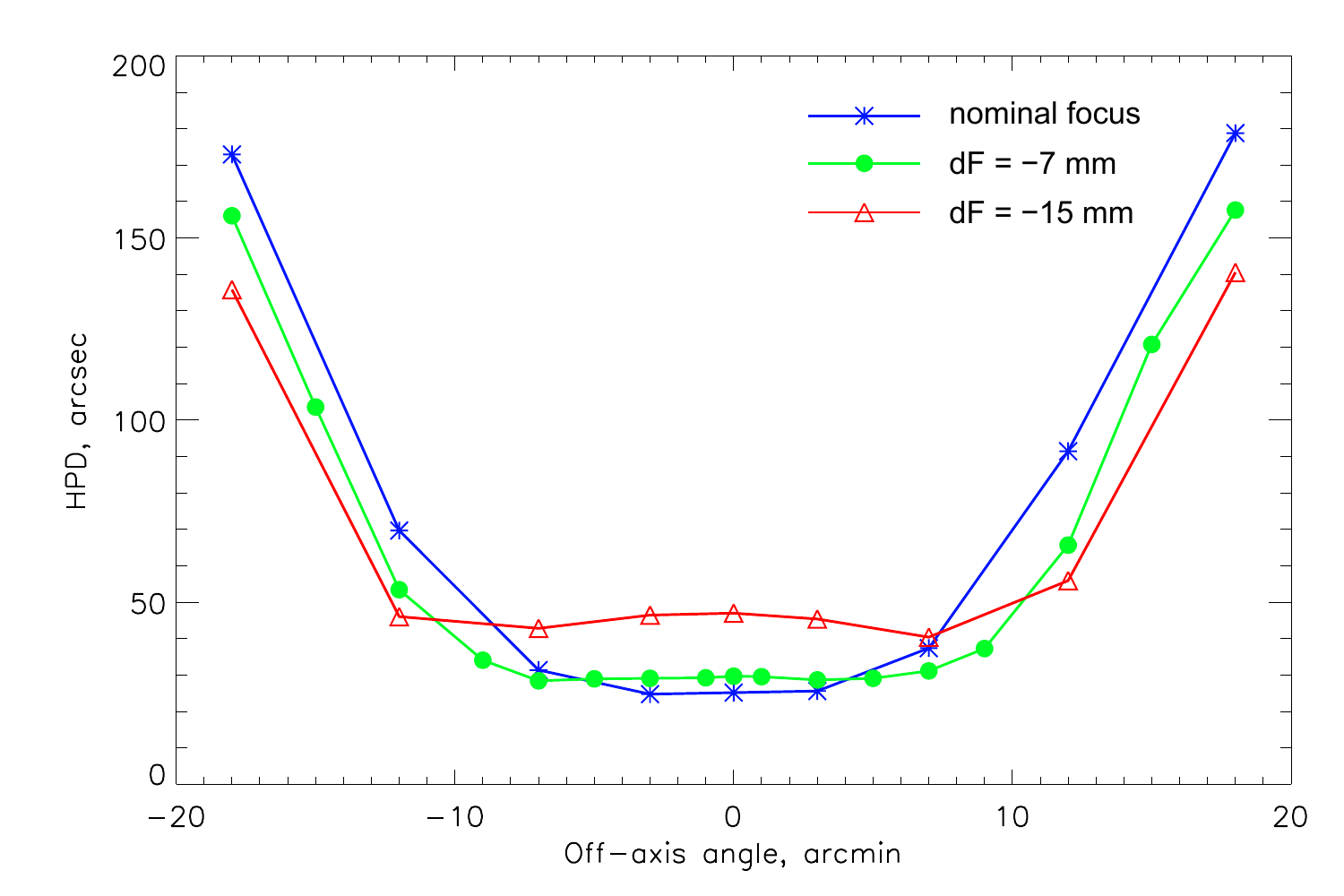}
}
\caption{The HPD of the mirror system for the nominal focus (2700~mm) and for two defocus positions of $-7$~mm and $-15$~mm. The HPD was measured at 8.1~keV for MS$_1$ at the MSFC Stray Light Test Facility.}
\label{fig:art_defocus}
\end{figure}

The weight of each \art\ MS is about 17 kg. The nominal focal length of the MSs is 2700~mm. During installation into the telescope (Fig.~\ref{fig:ms_sarov}), the MSs were defocused by 7~mm to provide a more uniform angular resolution across the field of view (see Fig.~\ref{fig:art_defocus}, \citealt{2014SPIE.9144E..4UG}). Thus, the distance from the MS to the detector plane is 2693~mm.

\subsection{X-ray detectors system}

The \art\ focal plane X-ray detectors \citep{2016SPIE.9905E..1JP,2016SPIE.9905E..51L} were developed in IKI specifically for the \srg\ mission.

The \art\ detector system consists of seven units of roentgen detectors (URDs with numbers from 01 to 07 according to the number of telescope modules), two blocks of electronics (BE) and one serial interface connection block (CB) \citep{2014SPIE.9144E..13L}. The power consumption of the detector assembly is 42~W, and the total mass budget is 39.6~kg. The two BEs consist of seven identical modules (four in BE01 and three in BE02). One BE module services one URD. Each module is connected to the supply distribution network and receives two pulse commands from the spacecraft control system. The BE module includes a primary power supply switcher, an electromagnetic interference filter, a set of low voltage DC-to-DC converters, a high voltage regulated DC-to-DC converter, and a current measure and protection circuit. The BE module provides all necessary low and high voltages to the URD and protect the latter from high current consumption. The CB is used for distribution of command/housekeeping and time synchronization interfaces. Both interfaces are based on the RS-485 standard with galvanic isolation.

\begin{figure}
\centerline{
\includegraphics[width=0.98\columnwidth,clip]{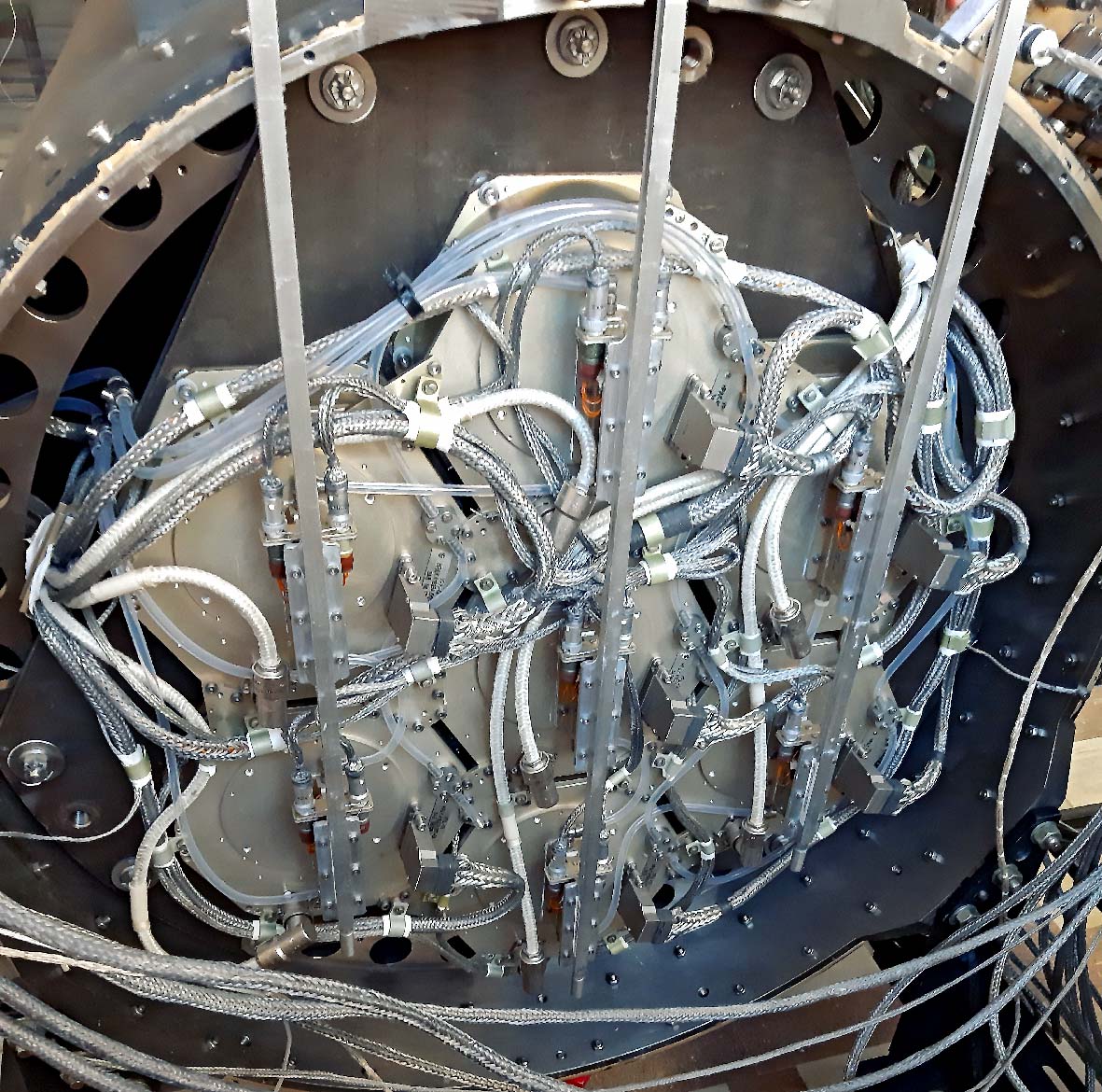}
}
\caption{Seven URDs in the \art\ telescope focal plane assembly (view of the back of the detectors).}
\label{fig:art_detect}
\end{figure}

The URDs and BEs are separated in order to lower the power dissipation inside the URD. One detection channel (one URD and one BE module) consumes 6~W from the on-board supply network. A power of 3.5~W is dissipated inside the URD and another 2.5~W is dissipated in the BE. The two BEs are placed on the thermally stabilized platform of the spacecraft. The seven URDs and CB are situated inside the \art\ telescope tube (Fig.~\ref{fig:art_detect}).

The URD performs the following tasks: receiving and processing commands from the subsystem storing on-board information (SSOI), processing time synchronization signals, providing all necessary biases to the detector, amplification and analog-to-digital conversion of the detector output signals, processing the digitized data, finding the areas with registered events, packing data to housekeeping frames, and transmitting housekeeping frames to the SSOI.

\begin{figure}
\centerline{
\includegraphics[width=0.98\columnwidth,clip]{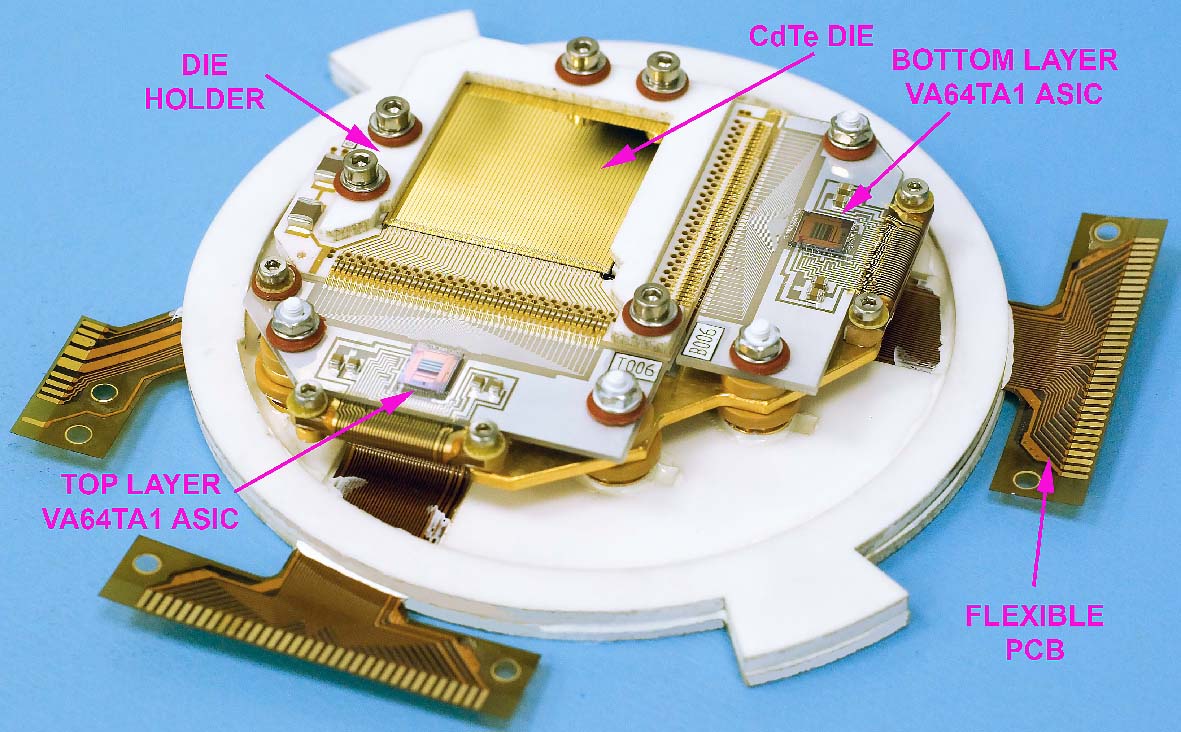}}
\caption{X-ray detector hybrid integrated circuit (model SD-01).}
\label{fig:art_detcirc}
\end{figure}

The X-ray detector, situated inside URD, is a big hybrid integrated circuit (Fig.\,\ref{fig:art_detcirc}). The sensitive elements of the detectors are high quality CdTe dies \citep{4774557}, manufactured by Acrorad (Japan). The size of the die is $29.95\times29.95\times1.00$~mm. The electrodes structure of the die is: (top) Au/Pt/CdTe/Al/Ti/Au (bottom). The structure CdTe/Al on the bottom electrode forms a Schottky barrier \citep{Toyama_2004}. A high-purity CdTe with the Schottky barrier provides a very low leakage current. To operate in a double side strip detector configuration, the top and bottom electrodes of the die are patterned by photolithography. On the top side, 48 parallel strips are formed, which are surrounded by a guard ring. The same pattern is formed on the bottom side, but it is rotated by $90\deg$ relative to the top side, which enables a reconstruction of two coordinates of an incoming event on the detector plane. The strips' width is $520$~$\mu$m, and the gaps between strips are $75$~$\mu$m. The sensitive area of the detector is $28.48\times28.48$~mm. Given the focal length of the MS ($\sim2700$ mm), the angular size of the strip is about 45$\arcsec$.

The strips of the die are connected to the inputs of two application specific integrated circuits (ASIC) VA64TA1 \citep{2006NIMPA.568..375T}, manufactured by Ideas (Norway). One ASIC services the top side, and the other services the bottom. The spectrometric channel of VA64TA1 consists of a charge sensitive amplifier, a fast CR-RC shaper, a discriminator, a slow CR-RC shaper, and a sample and hold unit. Each ASIC has 64 spectrometric channels, but only 48 are used. Due to some features of the VA64TA1 architecture, it has only one output trigger signal, and the determination of the triggered channel is performed by software. The URD can operate in two modes. In the first mode, it stores the time of an event as well as six amplitudes registered in three strips in the top layer and three strips in the bottom layer where the event was detected. This is the basic observation mode. In this mode, the URD produces one housekeeping frame (1024 bytes) per 81 events. In the second mode, the URD stores 96 amplitudes from all the strips in the top and bottom layers. In this mode, the URD produces one housekeeping frame (1024 bytes) per 3 events. Due to the high information content, the second mode is used for detector tests only.

The dead time of the detectors is less than 0.8~ms. The event registration time error relative to the spacecraft time synchronization signals is less than 23~$\mu$s.

\subsubsection{In-flight operation of the detectors }

During the flight, the thermal regime of the URDs and X-ray detectors is kept by the thermal control unit of \art. The operation temperature of the detectors is between $-22$ and $-19^\circ$C. The stability of the detectors' temperature is better than $\pm0.2^\circ$C per month.

The high voltage provided by the BE module can be regulated from 0 to $-360$~V, but the actually used operating voltage is $-100$~V. Such a relatively low biasing voltage is selected for three reasons: the \art\ working energy range is below 30~keV, to lower the leakage current of the die, and to derate the voltage regime of the components, servicing the high-voltage side of the CdTe die.

\begin{figure}
\centerline{
\includegraphics[width=0.98\columnwidth,clip]{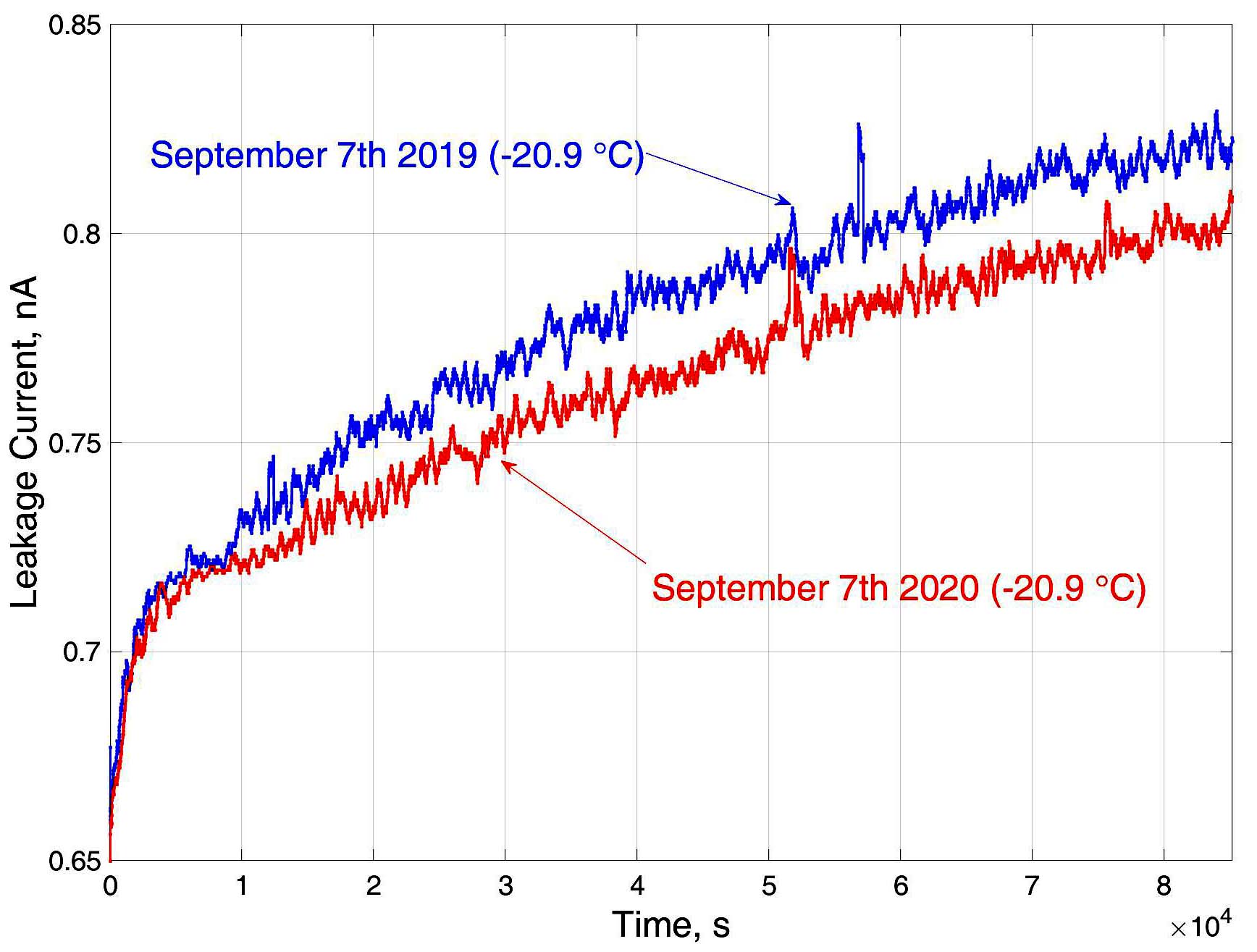}}
\caption{Leakage current vs. time for one-day operation of CdTe detector at $-100$~V.}
\label{fig:art_detinflt}
\end{figure}

Due to the well-known polarization effect in CdTe Schottky devices, we regularly switch off and on the high voltage applied to the CdTe die. At the commissioning stage, we tested all the detectors for two days of continuous operation and did not find any symptoms of polarization. Considering the low operation temperature of CdTe dies for a long time, we decided to make a depolarization once per day. The high voltage is applied to the detectors 23 hours 50 minutes per day and the detectors perform observations 23 hours 44 minutes per day. When the operating mode of one detector changes from observation to depolarization, all other detectors stay in observation mode. As soon as the depolarized detector returns to observation, we perform depolarization of the next detector and so on. The total time for depolarization of all the detectors is 1 hour 52 minutes. During operation, we monitor the leakage currents of all dies. We can measure only the total current through the volume of the die and guard rings together. One detector (in URD02) has the leakage current less than 850~pA, two detectors less than 400~pA and four detectors less than 100~pA. Figure~\ref{fig:art_detinflt} shows the dependence of the leakage currents of the detector in URD02 vs. time for one day of operation (zero is the time when high voltage switches on). After one year in flight, the leakage current has slightly reduced.

\subsection{Telescope electronics}

\begin{figure*}
\centerline{
\includegraphics[width=0.95\textwidth,clip]{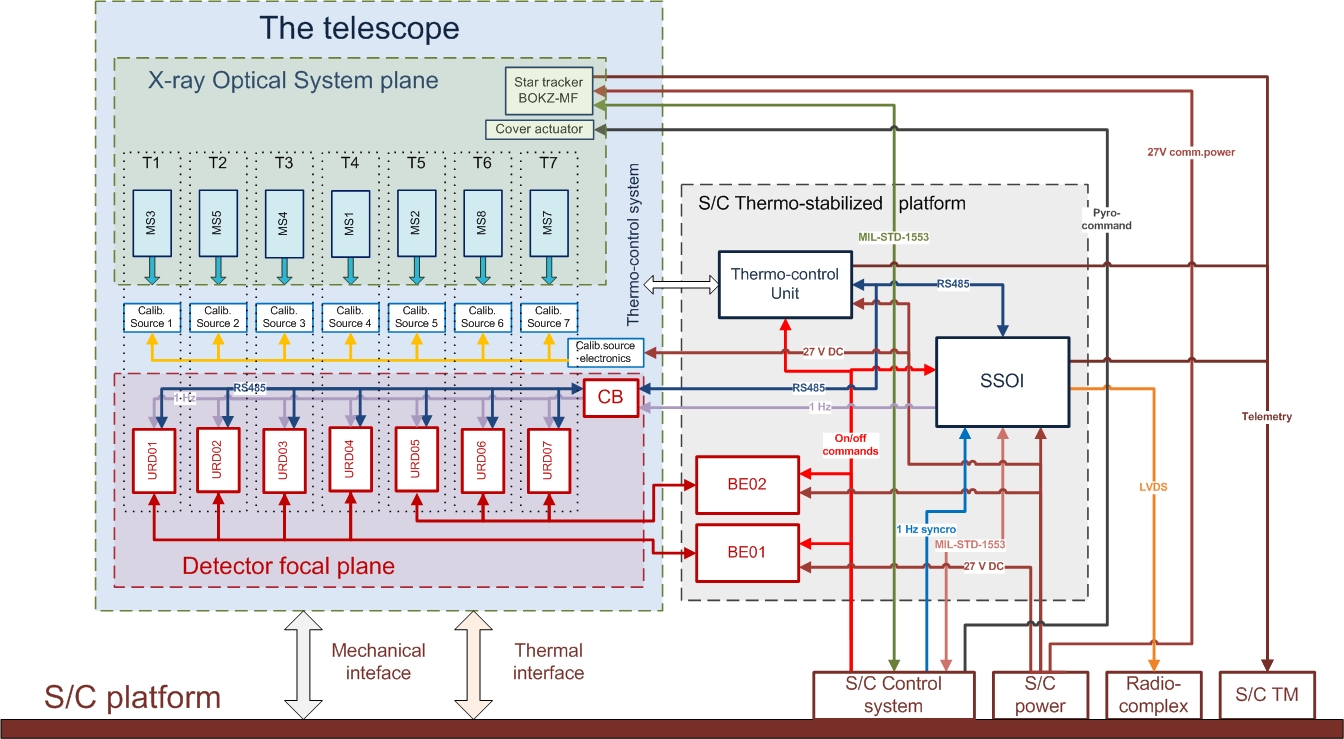}
}
\caption{Flowchart of the systems, units and interfaces of \art.}
\label{fig:art_interfaces}
\end{figure*}

The \art\ electronics includes several functional parts (see Fig.~\ref{fig:art_interfaces}):\\
\begin{itemize}
\item
X-ray detectors system;
\item
data processing unit (DPU, Russian abbreviation SSOI);
\item
star tracker (Russian abbreviation BOKZ-MF);
\item
thermal-control unit;
\item
calibration sources electronics.
\end{itemize}

SSOI continuously collects and stores data from the detectors, transmits data to the on-board radio system (radiocomplex) during ground contacts and provides electrical and logical interfaces with the S/C control system.

The thermal-control unit works continuously, controlling the electrical heaters of \art. As was mentioned above, there are 36 main and 36 redundant heaters in the telescope, and the same number of sensors. Each heater is controlled by its own sensor independently using a PID algorithm. Platinum Pt100 type temperature sensors are used.

All the electronic components have passed the necessary screening and additional tests, including radiation tests and destructive physical analysis.

\subsection{On-board data handling structure}

The data processing unit (SSOI) of \art\ performs:

\begin{itemize}
\item
data collection and storage from the seven detectors;
\item
receiving and storage of orientation data and quaternions from the star trackers and gyros;
\item
survey of the instrument's sensors (analog and digital) by the S/C telemetry system;
\item
synchronization of subsystems and data binding to the timeline;
\item
upload of telecommands for the instrument;
\item
download of scientific and housekeeping data.
\end{itemize}

\begin{figure*}
\centerline{
\includegraphics[width=0.8\textwidth,clip]{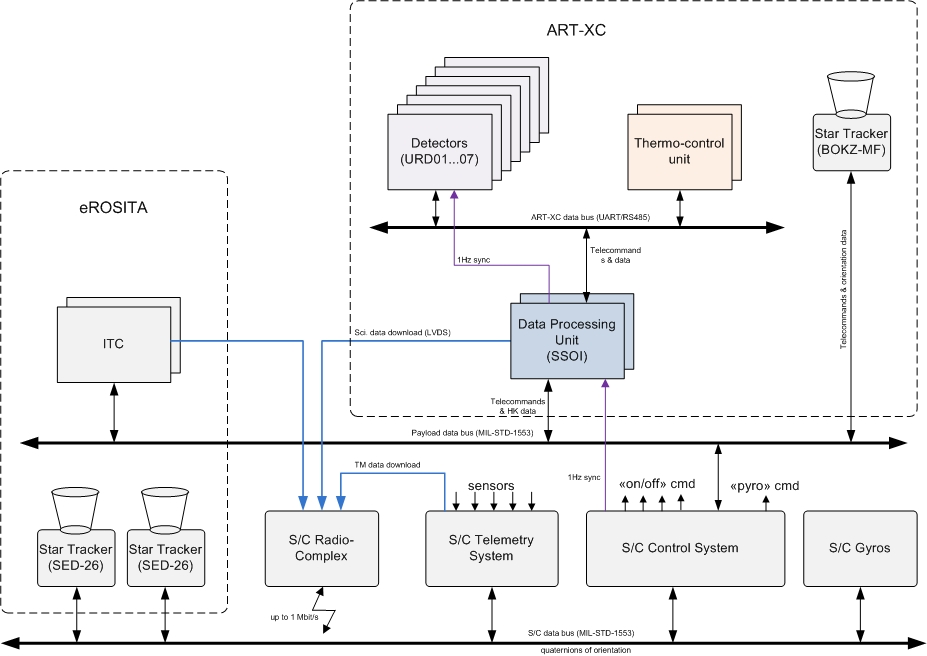}
}
\caption{The On-board Data Handling structure.}
\label{fig:art_obdh}
\end{figure*}

Scientific data recorded by the seven detectors and data from the thermal-control unit are transmitted via the \art\ data bus (UART/RS-485) to SSOI. The latter also receives orientation data and quaternions from the S/C gyros and star trackers (BOKZ-MF and SED-26), which are transmitted by the S/C control system (see Fig.~\ref{fig:art_obdh}). The typical volume of the \art\ data is 90--120~MB per day.

The FLASH based SSOI solid-state mass-memory has a capacity of 512~MB. This amount is enough to save scientific data for up to 5 days. Data is stored in memory using ECC Hamming codes, which makes it possible to correct single errors and detect double ones.

SSOI has a high-speed interface based on the LVDS (Low-Voltage Differential Signaling) physical standard with the on-board radiocomplex. The data structure of packets corresponds to CCSDS (Consultative Committee for Space Data Systems) 133.0-B-1. The radio system uses additional turbo-codes or convolutional codes for transmitting to the radio line. This ensures reliable data transmission during ground contacts.

Telecommands for operation of the instrument are sent to SSOI from the S/C control system (SCCS) by the payload data bus (MIL-STD-1553). Telecommands can be time-tagged (stored in the SCCS memory), deferred or directly executed. SSOI performs the function of transferring commands from the payload data bus (MIL-STD-1553) to the \art\ bus (UART/RS-485). The regular volume of commands for \art\ uploaded on board during ground contacts is about 1-2~Kbytes per day.

Data synchronization to the time line is implemented by transmitting the on-board time code (OTC) via information buses (MIL-STD-1553 and RS-485), as well as issuing 1~Hz synchro pulses (1PPS type) via the electrical interfaces to the detectors. The short term stability of the on-board clock generator is around $10^{-7}$, which results in a stable shifting of OTC of 10~ms per day. The OTC is adjusted periodically by special commands to S/C during ground contacts.

\subsection{Calibration X-ray sources}
\label{subsec:calsource}
\begin{figure}
\centerline{
\includegraphics[width=0.98\columnwidth,clip]{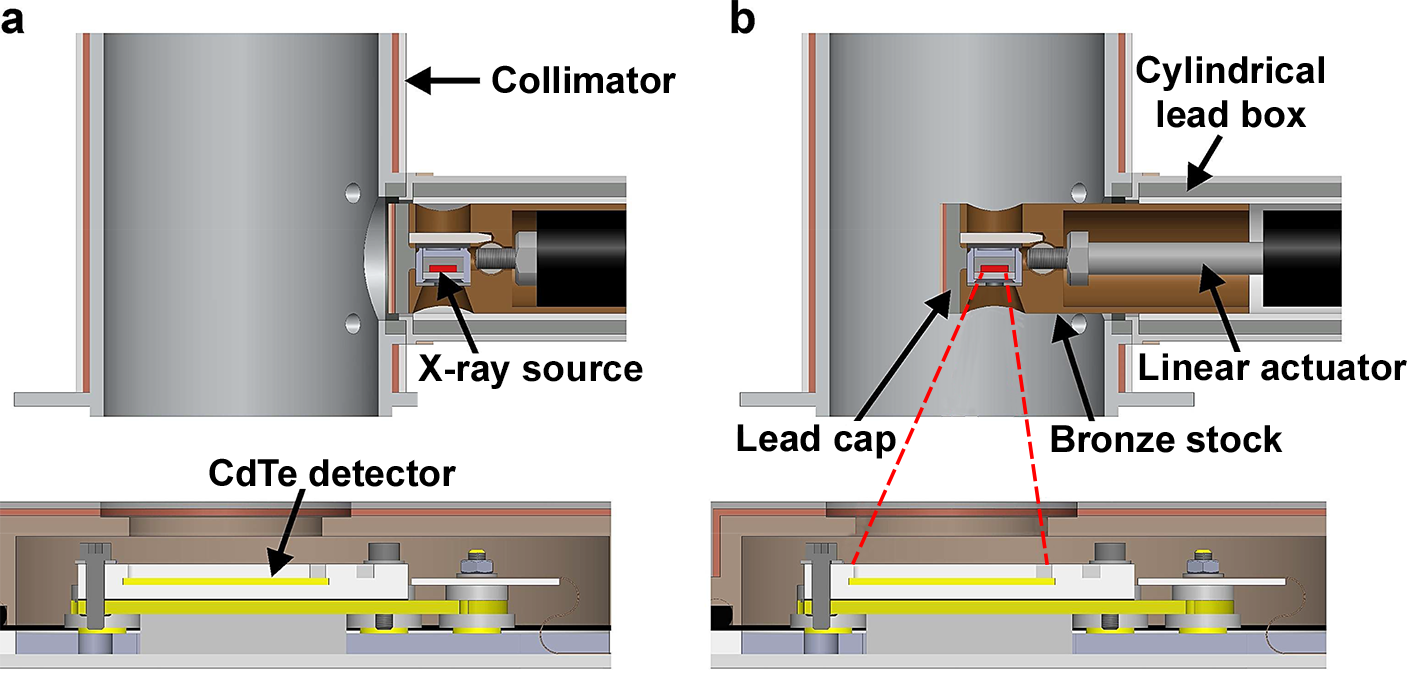}
}
\caption{A calibration source is stowed within a cylindrical lead box attached to each collimator of each telescope module (a). The lead box ensures no flux from the calibration source is measurable during astrophysical observations. The calibration source can be inserted into the collimator, as shown on the right panel (b), where it illuminates its detector.}
\label{fig:cal_src}
\end{figure}

All the detectors of the \art\ telescope are calibrated approximately every two months. For these purposes, the X-ray source ($^{241}$Am+$^{55}$Fe) moves into the field of view of the detector by means of the source calibration block, which includes a linear actuator. When the detector is observing, the calibration X-ray source is inside the lead box (Fig.~\ref{fig:cal_src}). To calibrate the detector, necessary commands are sent to the spacecraft. All the calibration sources work independently.

\subsection{Star tracker}

The \art\ star tracker BOKZ-MF is mounted on the MS platform next to the MSs and is aligned with the telescope’s optical axis. BOKZ-MF may be used as a redundant star tracker in the S/C attitude control system. The main parameters of BOKZ-MF are presented in Table~\ref{tab:bokz}.

There are also two SED-26 star trackers by SODERN aboard the spacecraft, whose attitude information is available to \art. These star trackers are mounted on the optical bench of \erosita\ and belong to the S/C control system.

\begin{table}[]
\caption{BOKZ-MF star tracker characteristics.}\label{tab:bokz}
\footnotesize{
\centerline{\begin{tabular}{l|l}
\hline
Parameter  &  Value \\
\hline
Mass & 2.8 kg \\
Power consumption & $\lesssim14$ W \\
Admissible S/C angular velocity & 1\,deg/s \\
Attitude data update frequency & 1 per second \\
Accuracy (rms) $\sigma$x,$\sigma$y/$\sigma$z & 2.5/25 \arcsec \\
Field of view & 14~deg \\
Stellar magnitude & up to 6 \\
Output data & Orientation matrix (cos) \\
Interface & MIL-STD-1553 \\
\hline
\end{tabular}}
}
\end{table}

\section{Ground calibrations}
\label{s:groundcal}

Ground calibrations of \art\ took place from 2014 to 2018, with three test facilities involved at different stages: NASA/MSFC's Stray Light Test Facility (USA), IKI's X-ray test facility (Russia), and the PANTER X-ray test facility (Germany).

\begin{figure}
\centerline{
\includegraphics[width=0.98\columnwidth,clip]{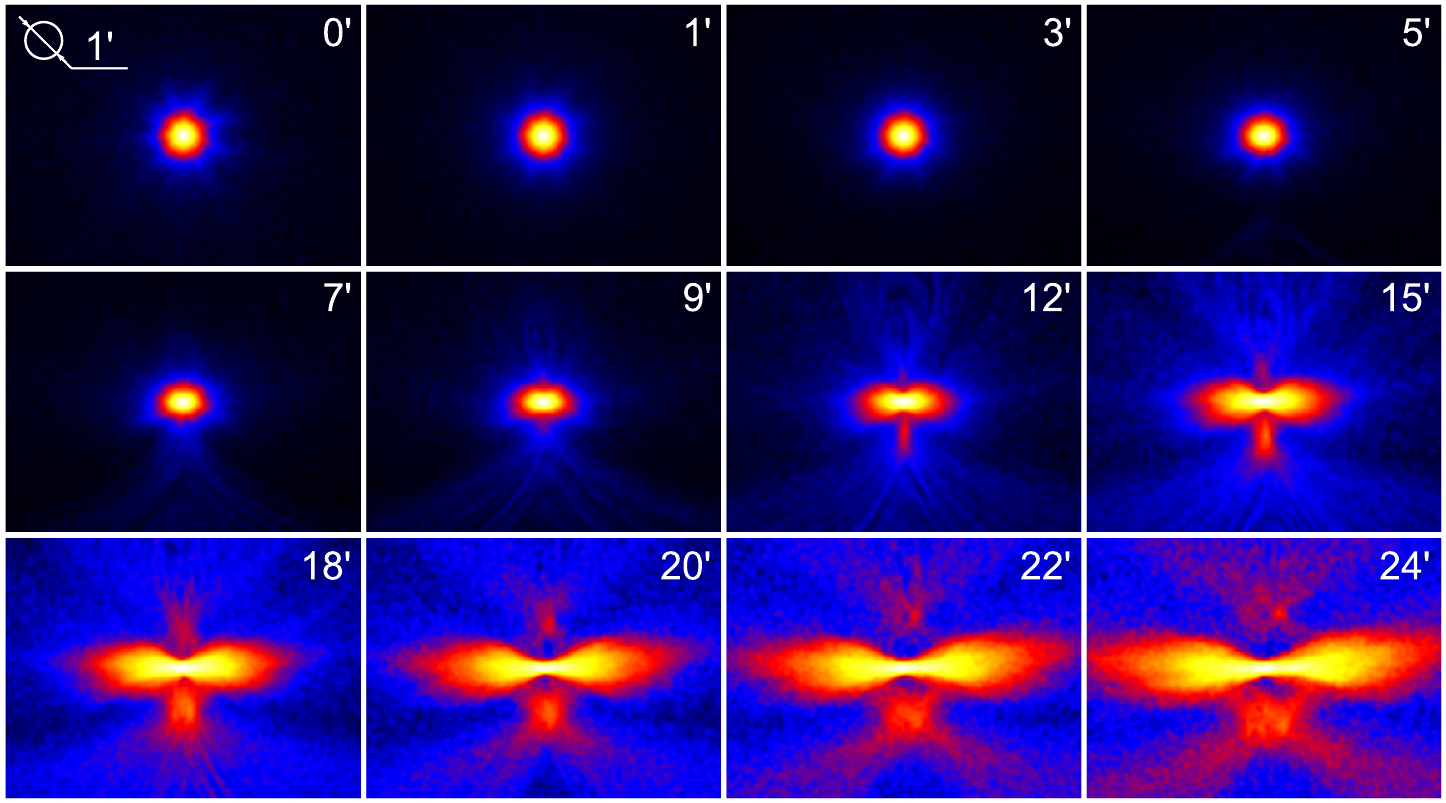}
}
\caption{Samples of the MS$_1$ PSF for different offset angles at 8.1~keV \citep{2017ExA....44..147K}. The PSF was obtained by averaging measurements at four azimuth angles. The size of each image is 400\arcsec$\times$560\arcsec. The color scale is logarithmic. The MS was defocused by 7~mm.}
\label{fig:psf_msfc}
\end{figure}

\begin{figure}
\centerline{
\includegraphics[width=0.98\columnwidth,clip]{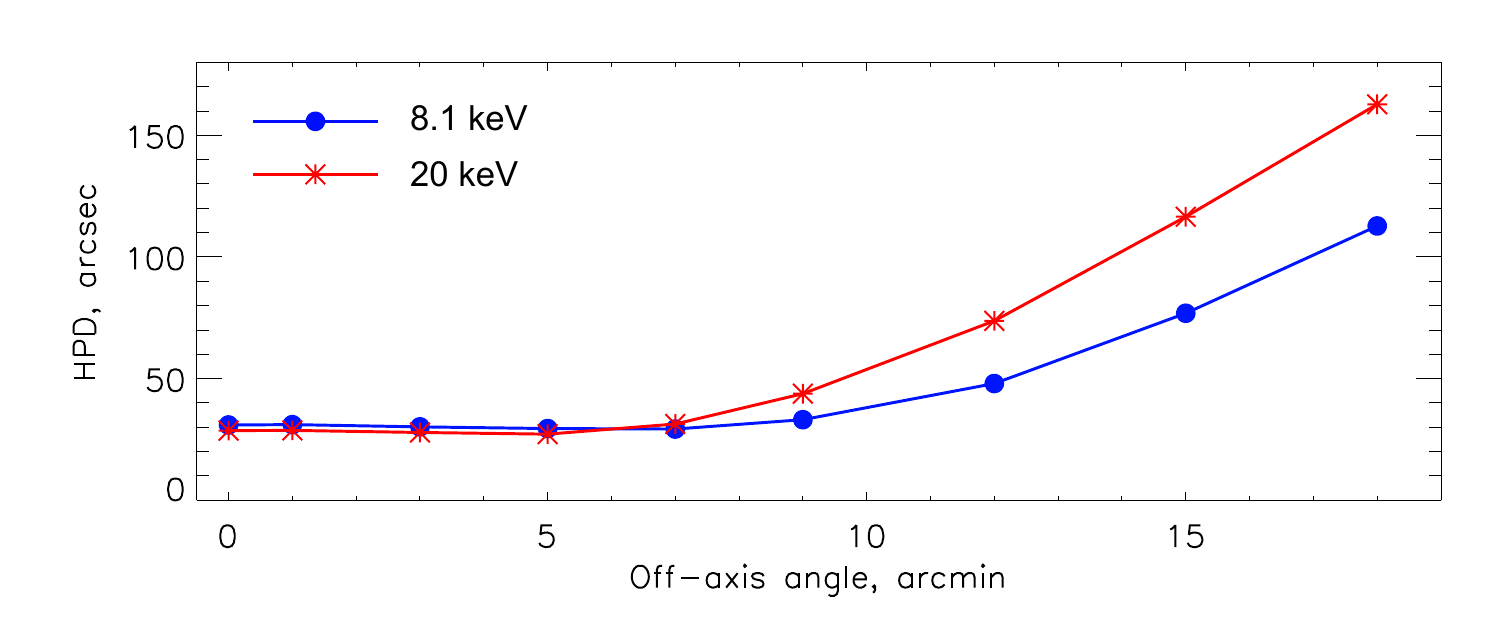}
}
\caption{The HPD of MS$_1$ at energies of 8.1 and 20~keV for several off-axis angles, as measured at MSFC's Stray Light Test Facility. The MS was defocused by 7~mm.  }
\label{fig:hpd_msfc}
\end{figure}

First, the \art\ flight MSs were calibrated at NASA/MSFC's Stray Light Test Facility \citep{2014SPIE.9144E..4UG,2017ExA....44..147K}. For each mirror system, the Half-Power Diameter (HPD), the Point-Spread Function (PSF) with high resolution, and the effective area at various off-axis angles and azimuths were measured. Figure~\ref{fig:psf_msfc} shows average PSF images obtained at different offset angles for one of the MSs. The HPD values at energies of 8.1 and 20~keV for several off-axis angles are shown in Fig.~\ref{fig:hpd_msfc}. Table~\ref{tab:artxc_mirrpar} lists the key derived parameters of each of the eight MSs at an energy of 8.1~keV. Seven of them have fairly similar characteristics and were installed in the \art\ flight model. One MS (MS6) showed somewhat poorer angular resolution and thus became the spare.

\begin{figure}
\centerline{
\includegraphics[width=0.6\columnwidth]{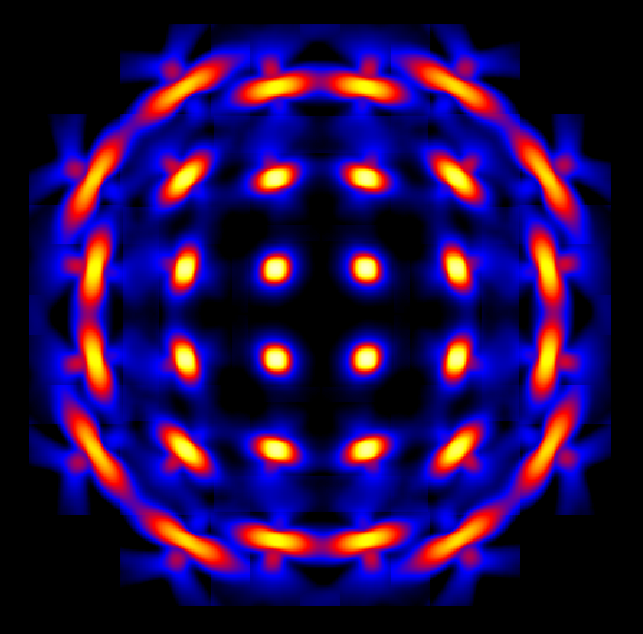}
}
\caption{\art\ PSF model applicable to data of survey and scanning observations, for various source positions in the telescope's field of view, at 8.1\,keV.}
\label{fig:psf_model}
\end{figure}

Based on these MS calibration data, an \art\ PSF model was constructed, which is used in application to \art\ data for X-ray source detection and characterization. Note that the effective \art\ PSF in survey and scanning modes is obtained through the convolution of the real MS PSF with a square window of the detector pixel size ($45^{\prime\prime}$). The resulting PSF model for various source positions in the telescope's field of view is shown in Fig.~\ref{fig:psf_model}.

\begin{table*}
\caption{Key parameters of the \art\ mirror systems at 8.1~keV.}\label{tab:artxc_mirrpar}
\centerline{\begin{tabular}{l|c|c|c|c|c|c|c|c}
\hline
System \# & MS$_1$ & MS$_2$ & MS$_3$ & MS$_4$ & MS$_5$ & MS$_6$ & MS$_7$ & MS$_8$ \\
\hline
HPD on axis, arcsec & 29.7 & 31.8 & 32.2 & 33.7 & 30.3 & 40.3 & 33.0 & 34.8 \\
W90 on axis, arcsec & 94.1 & 108.6 & 101.1 & 121.9 & 124.7 & 139.2 & 115.9 & 117.4 \\
On-axis effective area, cm$^2$ & 71 & 69 & 67 & 65.2 & 64 & 66 & 67 & 66.6 \\
\hline
\end{tabular}}
\end{table*}

As a next step, the \art\ flight MSs together with the detector units were calibrated at IKI's 60-m-long X-ray test facility in Moscow, Russia. For various reasons, the time allocated for the calibration of flight systems was severely limited. Therefore, additional, more detailed studies, were carried out using the spare mirror system (MS6) and one of the spare detector units (URD29) \citep{2018ExA....45..315P, 2019ExA....47....1P, 2019ExA....48..233P}. The main purpose of these tests was to determine the characteristics of the \art\ X-ray detectors, to verify the mathematical model of the \art\ mirror system, based on ray-tracing simulations, and to evaluate the telescope's performance.

\begin{figure}
\centerline{
\includegraphics[width=.98\columnwidth,clip]{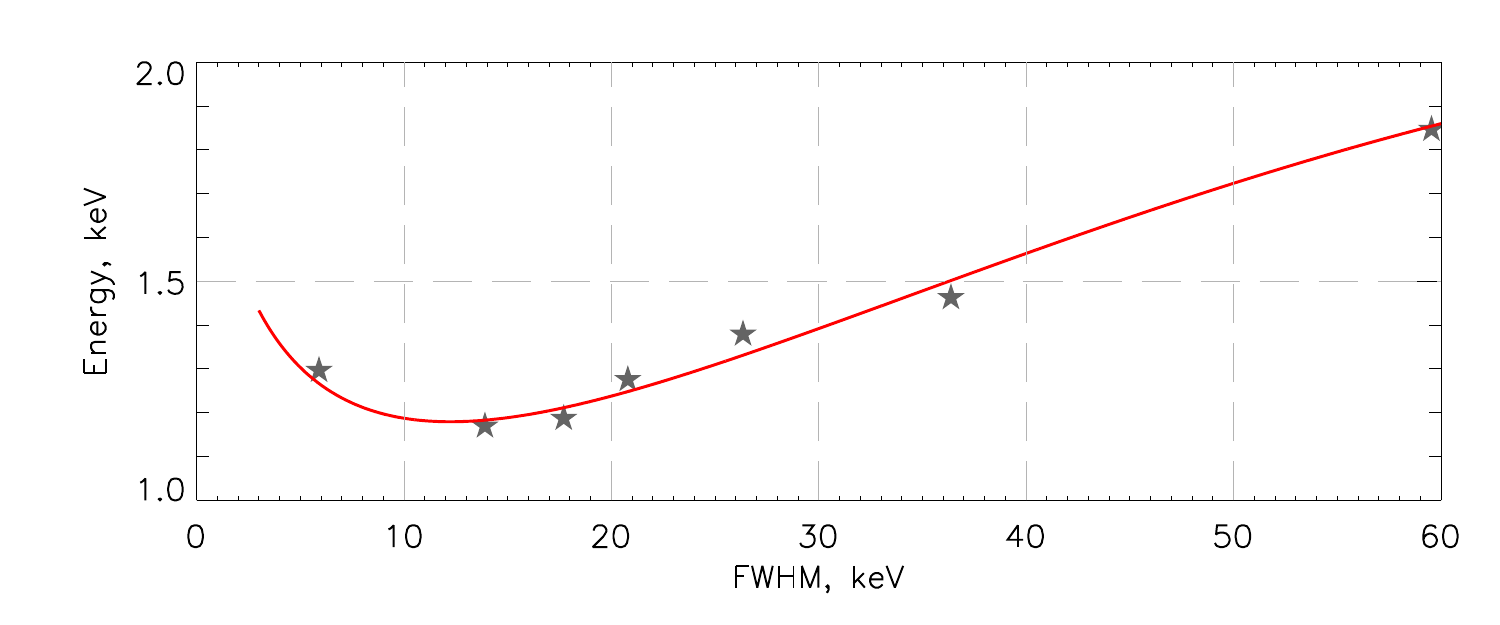}
}
\caption{Energy resolution (FWHM) of \art\ spare detector URD29 as a function of energy. The red curve is the approximation by a cubic polynomial function of $\rm{\sqrt{E}}$ in the 3--60~keV band.}\label{fig:det_spres}
\end{figure}

Figure~\ref{fig:det_spres} shows the measured energy resolution $\rm{\Delta E}$ (FWHM) of the \art\ spare detector unit URD29 as a function of energy. A significant deviation from the expected FWHM $\mysim\sqrt{E}$ dependence of energy resolution observed for URD29 at 5.9~keV is associated with a peculiarity of the ASIC VA64TA1 architecture. The ASIC circuit does not have a peak detector, which leads to a ``stretching'' of the peak shape towards low energies when processing signals with an amplitude close to the threshold.

\begin{figure}
\centerline{
\includegraphics[width=.98\columnwidth,clip]{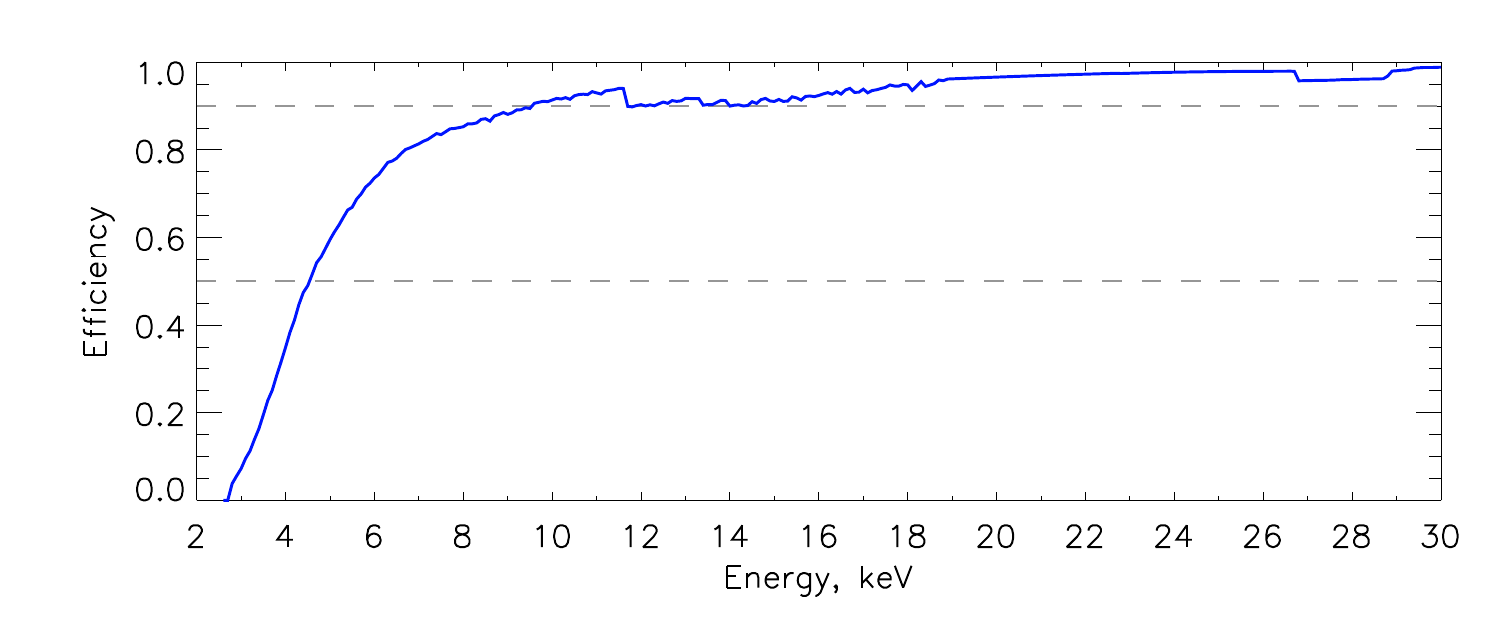}
}
\caption{Estimated efficiency of spare \art\ detector URD29 based on the model of the efficiency of a XR-100T-CdTe detector.}
\label{fig:urdeff}
\end{figure}

Figure~\ref{fig:urdeff} shows the efficiency of the \art\ spare detector, as measured at IKI's X-ray test facility. The efficiency reaches 50\% at 4.6~keV and exceeds 90\% at energies above 9.5~keV.

\begin{figure}
\centerline{
\includegraphics[width=0.98\columnwidth,clip]{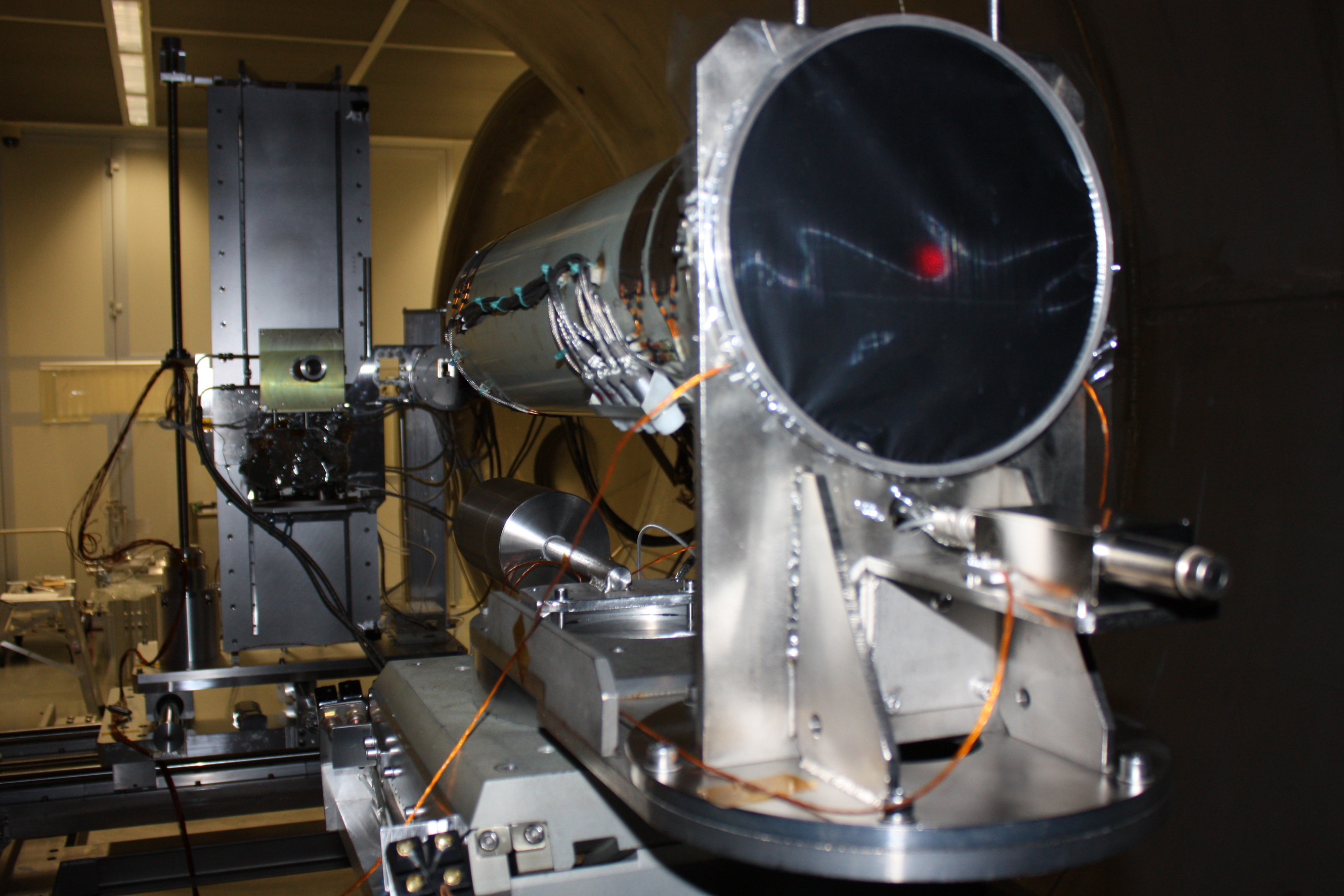}
}
\caption{Spare \art\ mirror system MS6 and detector unit URD29 installed at the PANTER X-ray test facility. }
\label{fig:panter_calib}
\end{figure}

The final stage of the \art\ ground calibration campaign took place at the PANTER X-ray test facility of the Max-Plank-Institute for Extraterrestrial Physics in Neuried, Germany in October 2018 (Fig.~\ref{fig:panter_calib}). For these tests, we used the \art\ spare mirror system and detector unit as well as the PANTER TroPIC pnCCD camera, which has also been used for ground calibrations of the \erosita\ mirror systems.

\begin{figure}
\centerline{
\includegraphics[width=0.98\columnwidth,clip]{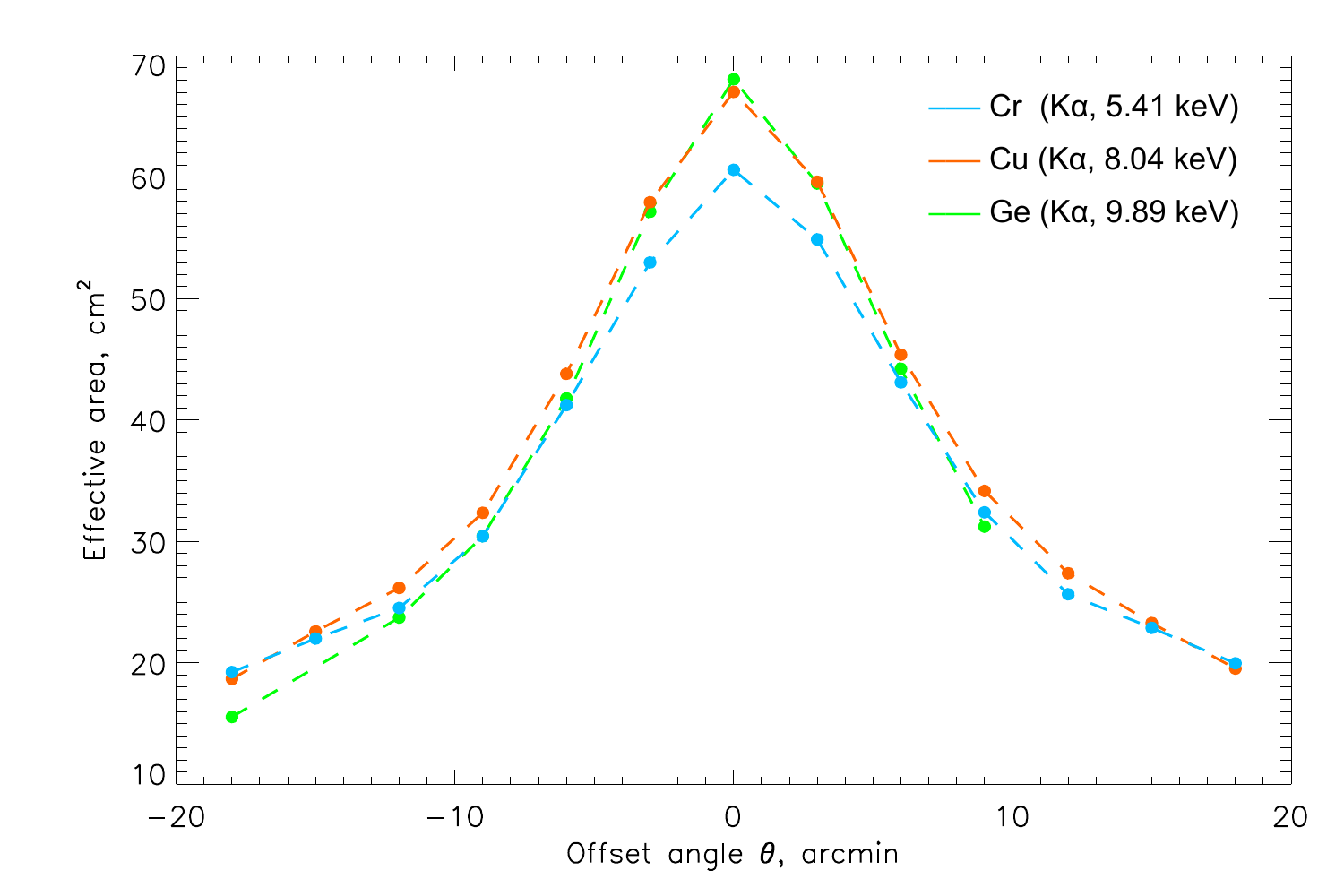}
}
\caption{Effective area of the \art\ spare mirror system as a function of the offset angle measured by the TRoPIC pnCCD camera at the PANTER X-ray test facility. The measurements were performed in the K$_\alpha$ lines of Cr (5.41~keV, blue points), Cu (8.04~keV, orange points) and Ge (9.89~keV, green points).}
\label{fig:panter_vign}
\end{figure}

During these calibrations, the on-axis effective area of the \art\ spare mirror was measured for the \art\ spare detector and for the TRoPIC camera. The results at the same energies are in good mutual agreement (see Table~\ref{tab:panter_effar}). Figure~\ref{fig:panter_vign} shows the effective area of the \art\ spare mirror system as a function of the offset angle measured by the TRoPIC pnCCD camera. The parameters of the on-axis PSF are given in Table~\ref{tab:panter_psfpar}.

\begin{table}
\caption{On-axis effective area of the \art\ spare mirror system measured at the PANTER X-ray test facility.}\label{tab:panter_effar}
\centerline{\begin{tabular}{c|c|c|c}
\hline
Source &Energy,& \multicolumn{2}{c}{Effective area, cm$^2$} \\
\cline{3-4}
 &keV& $\rm{URD_{29}}$ & TRoPIC \\
\hline
Ti & 4.51 & $60.4\pm0.6$ & --- \\
Cr & 5.41 & $60.9\pm0.4$ & $60.6\pm1.1$ \\
Cu & 8.04 & $66.8\pm0.4$ & $67.0\pm0.7$ \\
Ge & 9.89 & --- & $68.1\pm1.4$ \\
\hline
\end{tabular}}
\end{table}

\begin{table}
\caption{On-axis PSF parameters of the \art\ spare mirror system obtained using the TRoPIC pnCCD camera at the PANTER X-ray test facility.}\label{tab:panter_psfpar}
\centerline{\begin{tabular}{c|c|c|c}
\hline
Source & Energy, keV & HPD, arcsec & $W_{90}$, arcsec \\
\hline
Ti & 4.51 & $35.7$ & $159$ \\
Cr & 5.41 & $36.1$ & $152$ \\
Fe & 6.40 & $36.6$ & $170$ \\
Cu & 8.04 & $38.6$ & $225$ \\
Ge & 9.89 & $38.9$ & $241$ \\
\hline
\end{tabular}}
\end{table}

\begin{figure}
\centerline{
\includegraphics[width=.98\columnwidth,clip]{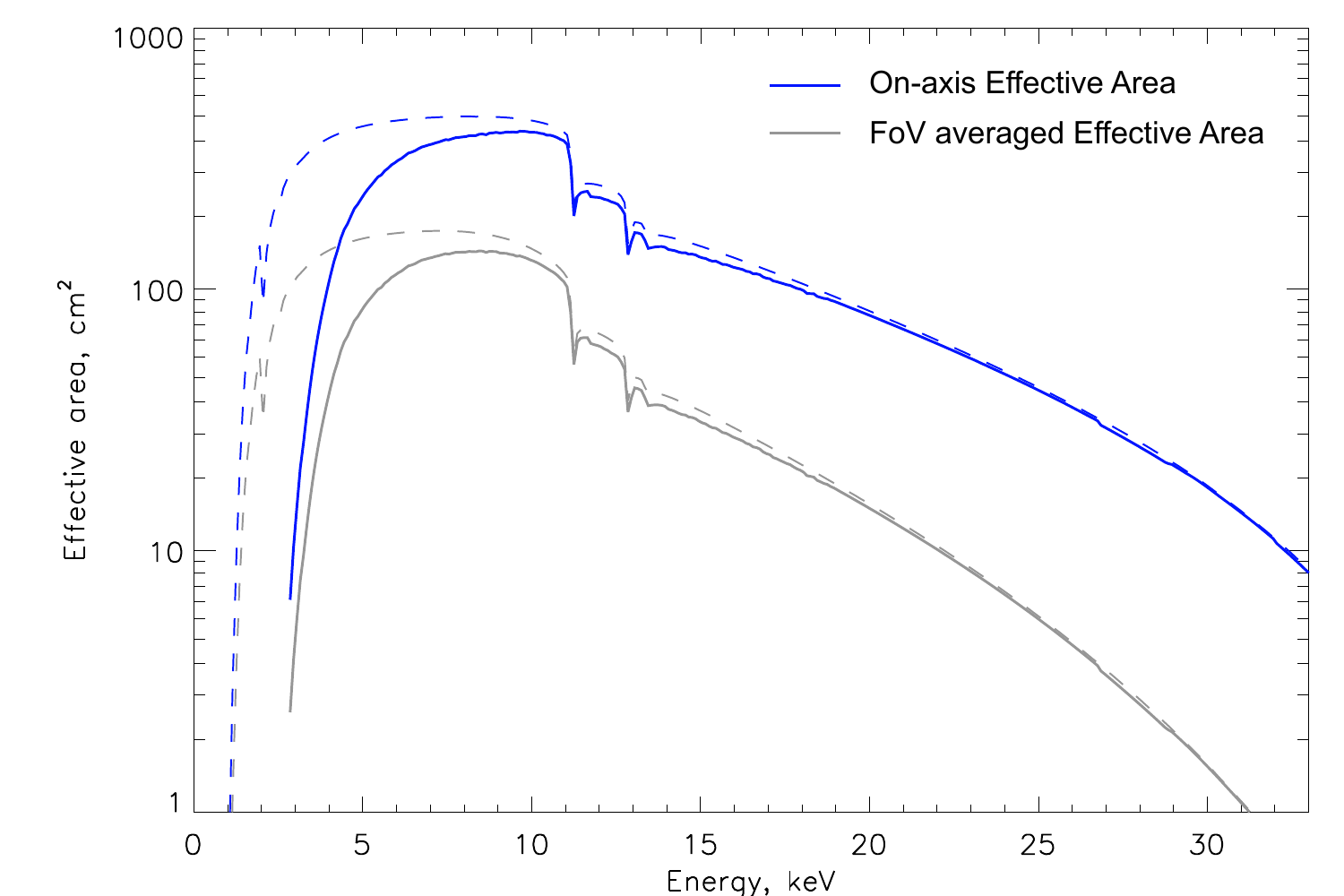}
}
\caption{\art\ on-axis effective area and FoV averaged effective area for doubly reflected events in the 4--35~keV energy band as a function of energy (solid curves). The \art\ effective area is based on the simulated mirror system's effective area (the dashed lines in this figure) and URD29 detector efficiency (Fig.~\ref{fig:urdeff})}
\label{fig:artxc_effarea}
\end{figure}

Assuming that all seven \art\ flight detector units have approximately the same efficiency as measured for spare detector URD29 and adopting the well-calibrated model for the effective area of the MSs, the on-axis effective area and grasp of the \art\ telescope have been estimated as a function of energy \citep{2019ExA....47....1P,2019ExA....48..233P}. Figure~\ref{fig:artxc_effarea} shows our estimates of the effective area of the \art\ telescope. The on-axis effective area at 8.1~keV is {$\mysim385$~cm$^2$}.

\section{Mission}

\subsection{Mission planning and timeline}

The \srg\ mission is operated by Lavochkin Association (also referred to as NPO Lavochkin or NPOL), the developer and creator of the Navigator platform. Mission planning is mainly done on a monthly basis. The plan of observations for the next month is prepared by IKI and checked by NPOL for possible observational constraints. The approved month observational program is normally divided into schedule blocks, which can be sent to the spacecraft on a daily basis during the ground contacts. This naturally sets the constraints for the \srg\ response time for possible transient events from one day to several days. The actual \srg\ observing schedule is available on a dedicated website \url{http://srg.cosmos.ru}.

\begin{figure}
\centerline{
\includegraphics[width=0.98\columnwidth,clip]{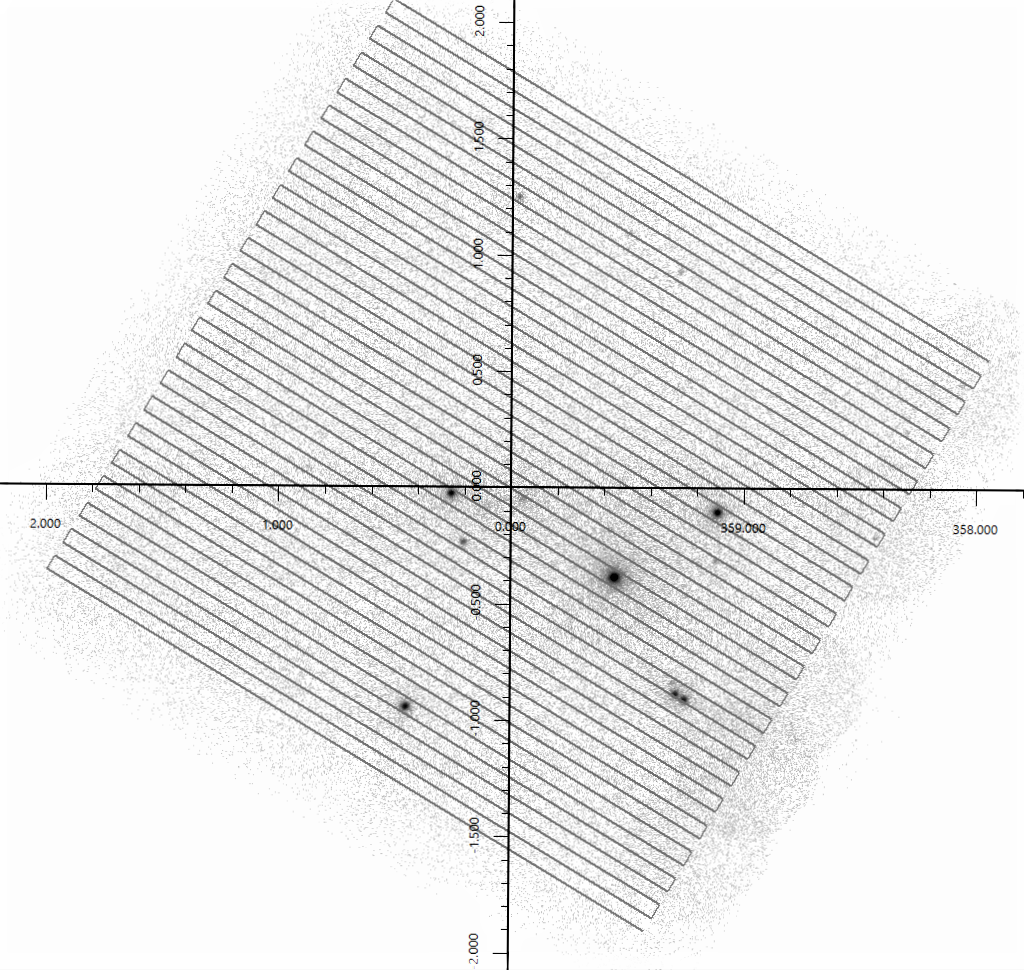}
}
\caption{Example of a scanning mode observation: Galactic Center scan on 10 Sept. 2019 (Galactic coordinates).}
\label{fig:art_scan}
\end{figure}

Table~\ref{tab:timeline} summarizes major milestones for \art\ from the \srg\ launch to the beginning of the all-sky survey. \\

\begin{table*}
\caption{The main events of the \art\ flight from launch to the start of the survey. }\label{tab:timeline}
\begin{tabular}{l|l|l}
\hline
Date & Time & Event \\
\hline
2019/07/13 & 15:31 & Launch \\
2019/07/13 & 17:31 & End of s/c acceleration and separation of DM-03 booster from s/c \\
2019/07/13 & 18:43 & Telescope subsystem switched on and stores on-board information \\
2019/07/13 & 18:58 & Telescope thermal control system switched on \\
2019/07/18 & 18:42 & First telemetry from first X-ray camera received \\
2019/07/21 & & All cameras switched on and reported “healthy” \\
2019/07/22 & & First trajectory correction \\
2019/07/23 & 18:31 & Telescope cover opened\\
2019/07/27 & & First X-ray calibration source opened and first calibration of first camera performed, \\ & & start of cameras commissioning \\
2019/07/30 & 17:29 & "First Light" (image of Cen X-3 on all cameras) \\
2019/08/06 & & Second trajectory correction \\
2019/08/25 & & Finish of seven cameras commissioning, EEPROM of first camera successfully reprogrammed \\
& & to working configuration, start of \art\ calibration and performance verification phase \\
2019/10/05 & & Finish of the \art\ calibration and performance verification phase \\
2019/10/21 & & Third trajectory correction \\
2019/12/12 & & Start of the all-sky survey \\
\hline
\end{tabular}
\end{table*}

\subsection{\art\ Observing Modes}
\label{sec:scans}

There are three modes of observations with \art\ aboard the spacecraft: (1) pointed observations mode, (2) survey mode, and (3) scan mode.

In the pointed observations mode, the optical axis of the telescope is fixed in a given direction. This mode is usually used for calibration observations.

The survey mode is used during the all-sky survey. In this mode, the telescope’s optical axis is rotated with a period of 4 hours around the spacecraft axis, pointed towards the Sun. This enables the full sky coverage in about 6 months. We can control movement parameters of the rotation axis by setting the plane of rotation, rotation speed and initial direction. Typically this is done once per week. During the first two surveys the rotation plane coincided with the Ecliptic plane.

The scan mode is the third observing mode realized in the \srg\ mission, which enables observations of fairly large regions of the sky ($12.5\deg \times 12.5\deg$ maximum) with uniform exposure. In this mode, the optical axis of \art\ is performing a “snake scan”. The spacecraft control system is automatically conducting a series of repeated consecutive rotations around two spacecraft axes with a set of predefined parameters, this set is being called a ''template''. Before being used this ''template'' has to be checked by and agreed with the spacecraft control team.

The scan mode was widely used during the Calibration and Performance Verification phase as well as during additional calibration sessions, and provided excellent results (see Fig.~\ref{fig:art_scan}).

\subsection{Science operation center}

\begin{figure}
\centerline{
\includegraphics[width=0.8\columnwidth,clip]{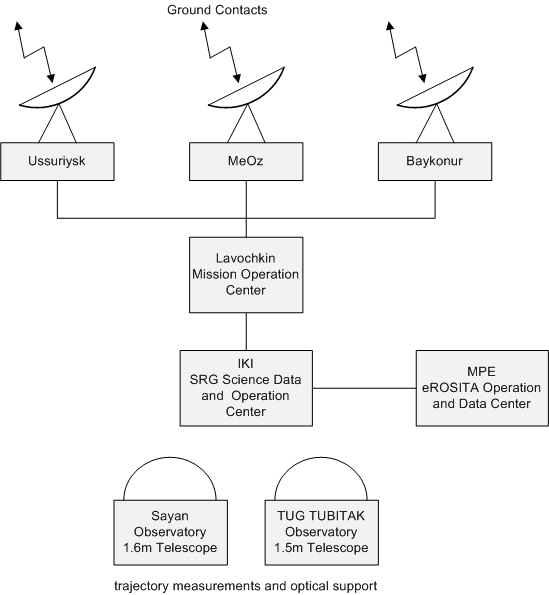}
}
\caption{The ground science segment.}
\label{fig:art_Gcontact}
\end{figure}

The ground science segment is shown schematically in Fig.~\ref{fig:art_Gcontact}. The main objectives of the science operation center are the hardware state control, health monitoring and telemetry dumping, as well as unpacking, verification and primary processing of the science data. These operations are performed on a daily basis during the mission ground contacts. A typical daily routine includes the following stages:

\begin{itemize}
	\item monitoring of the telescope hardware, including various thermal, currents and voltages sensors data, statuses of on-board hardware, amount of the data available, etc;
	\item mass memory dumping. The data volume is on average $\sim 100$ Mb/day in survey mode. All telemetry is transferred to the ground data receiving stations, then to the Mission operation center and further to the Science operation center;
	\item unpacking, verifying integrity and processing the data (including making of quicklook products);
	\item analysing the detailed scientific hardware statuses;
	\item analysing the quicklook report for anomalies present in the data;
	\item holding real-time command control sessions (only in emergency cases, to be able to reconfigure hardware instantly during ground contact);
	\item scheduling the next day operation program.
\end{itemize}
The \art\ ground segment operations take on average $\sim 20 \%$ of the total ground contact time.

\subsection{Science data center}

All \srg/\art\ telescope science analysis and data archiving are hosted within the IKI science data center. It includes: data reduction, near real-time analysis, sky survey processing, optical ground support and data archiving. Additionally the IKI science data center is responsible for the same tasks for the \erosita\ telescope in the part of the Russian quota of the scientific data (Sunyaev et al., 2020).

\subsubsection{Data reduction} \label{sssec:data_reduction}

The \art\ data reduction software provides tools for the creation of clean calibrated science products.

\begin{figure}
\centerline{
\includegraphics[width=0.98\columnwidth,clip]{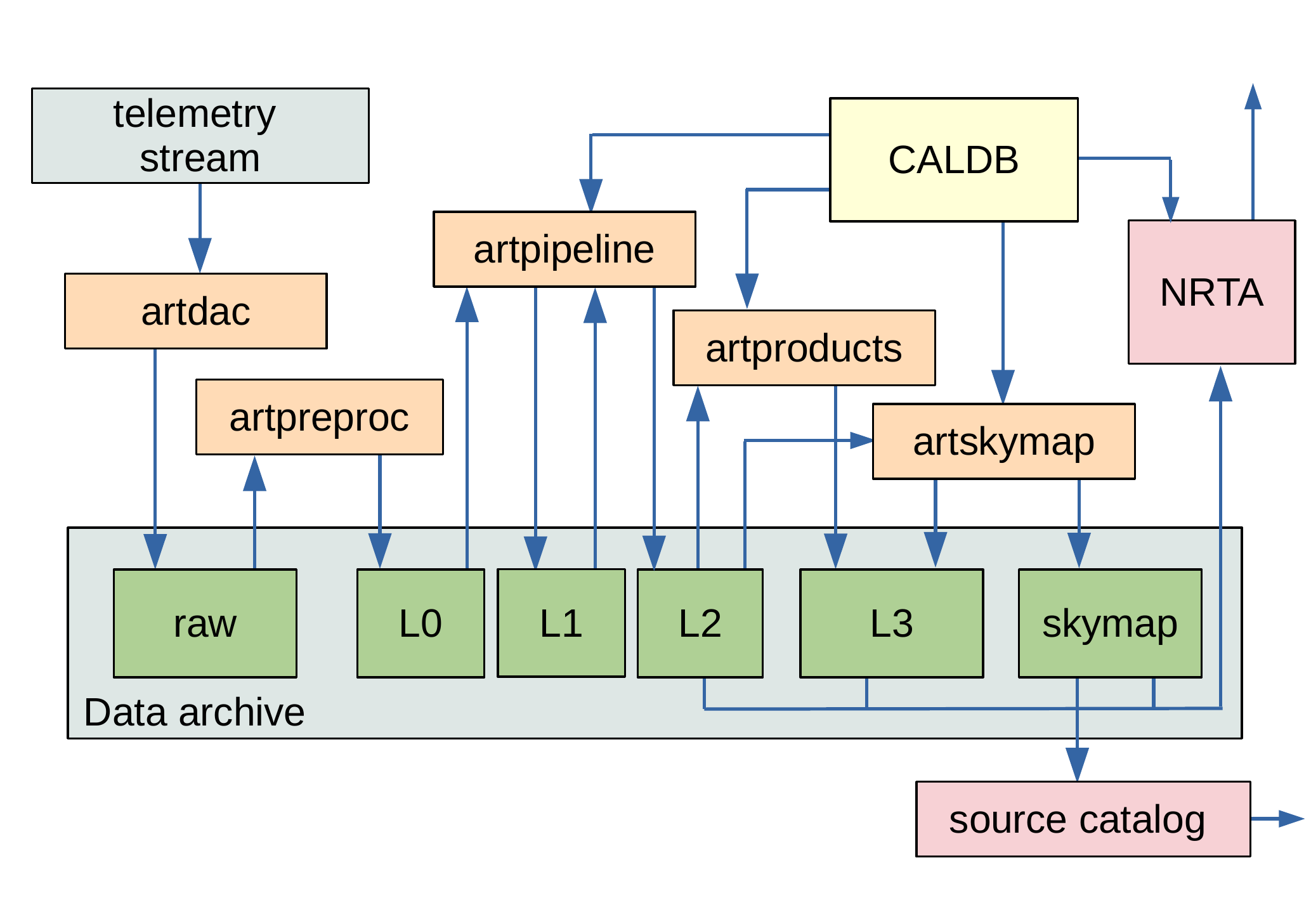}
}
\caption{\art\ data processing scheme.}
\label{fig:artpipeline}
\end{figure}

The software is organized as a number of independent command line tasks (Fig.~\ref{fig:artpipeline}). These tasks can be chained in sequences (using a control task or manually) for producing science data of different levels and extracting various scientific products. The data processing scheme is shown in Fig. \ref{fig:artpipeline}.

The data processing levels:
  \begin{itemize}
    \item \textsc{raw} -- raw telemetry data (binary formats)
    \item \textsc{l0} -- telemetry data (FITS)
    \item \textsc{l1} -- calibrated science data (FITS)
    \item \textsc{l2} -- cleaned science data (FITS)
    \item \textsc{l3} -- science products (images, light-curves, spectra, etc.)
    \item \textsc{skymap} -- L2 and L3 all-sky survey data in sky tiles, including images, exposure and background maps, X-ray source catalogues, etc.
  \end{itemize}

The telemetry stream received during a daily ground contact is verified for transmission errors and separated by detector and star tracker units using the \textsc{artdac} tasks and tools. This \textsc{raw} data is then processed by a preprocessor (\textsc{artpreproc}), which reads the \art\ telemetry, verifies its integrity and performs conversion into a standard FITS format. The \textsc{l0} data is processed by a pipeline software (\textsc{artpipeline}), which at first produces calibrated data products (\textsc{l1}) and then cleaned science data (\textsc{l2}). Calibrated data products includes event lists for each of the telescopes and a spacecraft attitude data. For each event in the event lists a correct photon energy and celestial coordinates is calculated, time of arrival corrections is applied and finally an event grade is assigned. A spacecraft attitude reconstruction is performed using orientation data available (BOKZ-MF, SED-26 star trackers and an inertial navigation system GYRO as well). Cleaned science data is suitable to be used in scientific analysis and is a result of a screening of event files by applying various criteria to attitude/orbital/instrument parameters and event properties.

The cleaned science data are processed further (\textsc{artproducts}) to obtain science products (\textsc{l3}), including images, light-curves, spectra, etc. The all-sky survey data are organized in 4,700 overlapping $3.6^\circ\times3.6^\circ$ sky tiles (\textsc{artskymap}), which are updated on daily basis to produce \textsc{l3} data products, including X-ray source catalogs (see below). The majority of science data products at \textsc{l3} are standard-compliant FITS files and may be used in standard X-ray analysis software (e.g., \textsc{heasoft}).

The sky survey products are processed and organized using an all-sky map of 4,700 sky tiles. The data in tiles is updated on a daily basis following the appearance of science data products. A set of standard L2 and L3 products is prepared for each sky tile, including exposure map, particle and photon background maps, and sky images. The appearance of sky survey products triggers a source detection procedure, which produces a source candidates catalog. Source detection and characterization is done by wavelet decomposition (\texttt{wvdecomp}, \citealt{1998ApJ...502..558V}), matched filter and maximum likelihood fitting algorithms. These use a complete PSF model as a function of source offset angle, constructed using the results of \art\ MSs calibrations at MSFC (see \S\ref{s:groundcal}). The source lists undergo cross-correlation and source identification procedures using various existing X-ray source catalogs and astronomy databases. The resulting sky survey data, source catalogs and sky tile images are made accessible for authorized users through a dedicated web interface.

The end-user software will be made public along with the opening of the \art\ data. Details of the data reduction scheme and software will be discussed in future publications.

\subsubsection{Near real-time analysis}

The science data products obtained during the data reduction stage are analysed further by the automatic near-real-time analysis (\textsc{nrta}) system. The system performs computation of the count rate (in the 4--12~keV energy band), produces background and exposure maps of the sky segment covered during the previous day. A simple sliding window source detection algorithm is then applied to the science data products. Finally, the resulting \art\ daily source candidates catalog is cross-correlated with X-ray and general source catalogs available in the system and reported to the \art\ transients group on a time scale of about an hour.

\subsubsection{Ground support at optical telescopes}

Optical telescopes included in the ground segment solve two important tasks. The first one is the trajectory measurements of the \srg\ spacecraft with the purpose to know better its position, that is especially important before and after the orbit corrections. Another task is the observations of potentially interesting astrophysical objects uncovered in X-rays by \art\ in order to identify their nature, measure their redshifts, etc. Such observations are, in particular, carried out at two dedicated optical telescopes, which are part of the \srg\ mission ground segment: the Sayan observatory 1.6-m telescope (AZT-33IK), operated by the Institute of Solar-Terrestrial Physics of the Siberian branch of the Russian Academy of Sciences, and the Russian-Turkish 1.5-m telescope (RTT-150), operated jointly by the Kazan Federal University, the Space Research Institute (IKI, Moscow), and the TUBITAK National Observatory (TUG, Turkey).

\begin{figure}
\centerline{
\includegraphics[width=0.98\columnwidth,bb=90 170 570 393, clip]{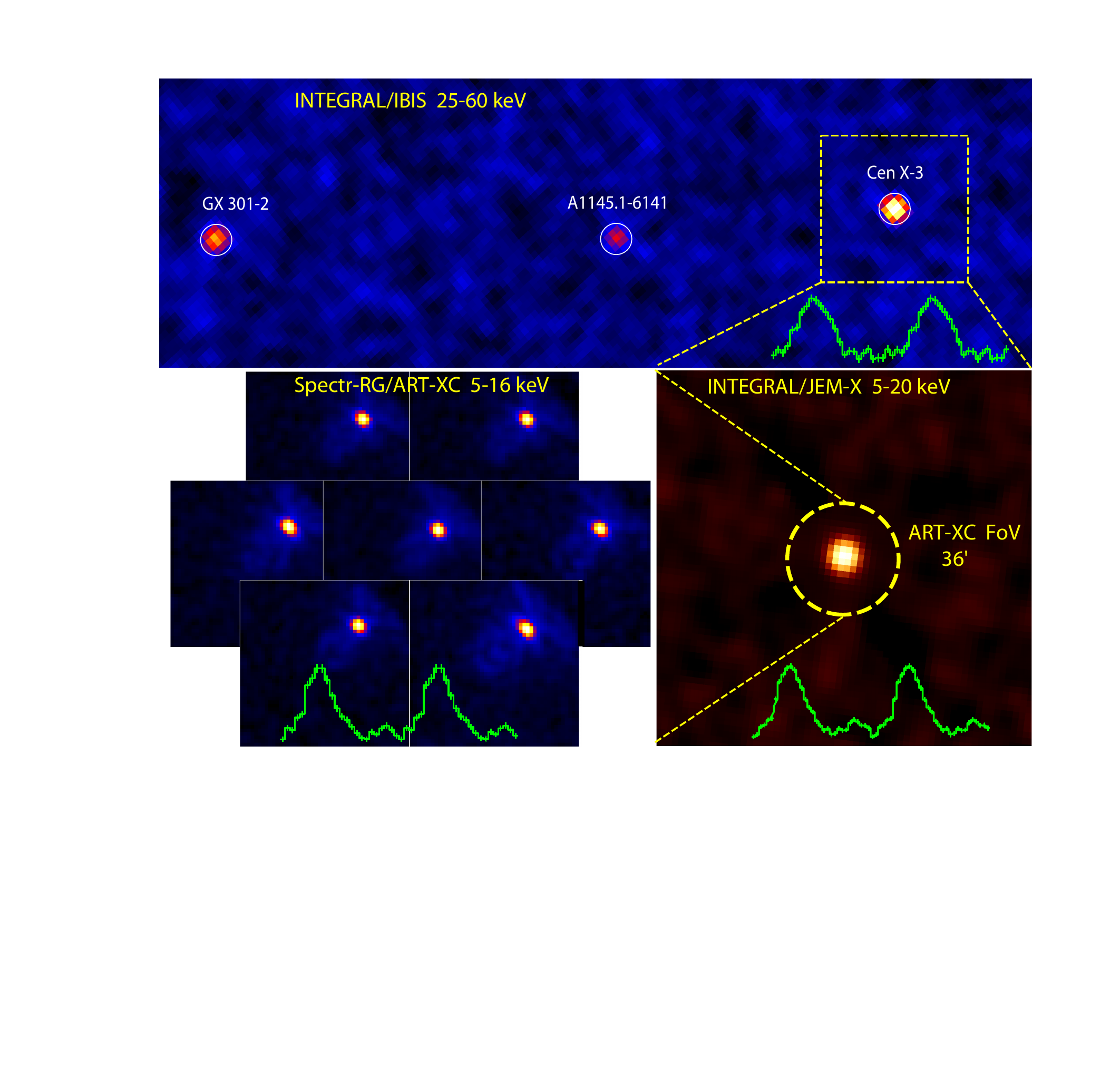}
}
\caption{Images of Cen\,X-3 obtained by the seven \art\ modules in the 5--16~keV energy band (left) simultaneously with {\it INTEGRAL}/JEM-X (right). The phase curves of the pulsating signal with the period of 4.8~s are in good agreement for both instruments (green curves on each panel).
}
\label{fig:art_int}
\end{figure}

\section{Scientific goals and expectations}
\label{s:goals}

The main goal of \art\ is to survey the whole sky in the medium X-ray energy range of 4--12~keV with the record-high sensitivity of $\sim 10^{-12}$~\flux\ ($\sim 10^{-13}$~\flux\ near the ecliptic poles) and sub-arcminute angular resolution.
The relatively hard energy band of \art\ is particularly well suited for studying populations of astrophysical objects that are significantly affected by intrinsic X-ray photoabsorption, such as active galactic nuclei (AGN), high-mass X-ray binaries (HMXBs) and cataclysmic variables (CVs).

Over the last two decades, a number of all-sky (serendipitous) surveys have been performed in energy bands close to that of \art. The \textit{RXTE} Slew Survey (XSS) achieved a sensitivity $\sim 10^{-11}$~\flux\ and angular resolution $\sim 1^\circ$ in the 3--20~keV energy band in the extragalactic sky ($|b|>10^\circ$) \citep{2004A&A...418..927R}. A somewhat better sensitivity ($\sim 5\times 10^{-12}$~\flux) has been reached in the 4--10~keV energy band in the \textit{MAXI}/GSC all-sky survey \citep{2018ApJS..235....7H,2018ApJS..238...32K}, which similarly to XSS is conducted by a collimator instrument. A similar average depth in the 2--12~keV energy band, but with excellent angular resolution (thanks to the use of mirror X-ray optics), characterizes the \textit{XMM-Newton} Slew Survey (XMMSL) \citep{2008A&A...480..611S}; however, XMMSL has covered the sky very non-uniformly and not in its entirety (84\%). In addition, there are hard X-ray (above 15~keV) all-sky (also serendipitous) surveys carried out by the coded-mask \textit{INTEGRAL}/IBIS and \textit{Swift}/BAT instruments, which have reached a depth $\sim 10^{-11}$~\flux\ at angular resolution $\sim 5$--$10$~arcmin (e.g. \citealt{2007A&A...475..775K,2012A&A...545A..27K,2021NewAR..9201612K,2010A&A...524A..64C,2007ApJS..170..175B,2016ApJS..223...15B,2016MNRAS.459..140M,2018ApJS..235....4O}).

The \art\ all-sky survey will significantly improve on all of these surveys in terms of the combination of angular resolution, sensitivity and uniformity, and will provide a rich astrophysical database for explorations of Galactic and extragalactic objects, as briefly outlined below.

\subsection{Extragalactic objects}

\subsubsection{Active galactic nuclei}

Because X-ray source detection is much less affected by photoabsorption in intervening gas in the 4--12~keV energy band compared to softer bands, the \art\ survey will provide unique data for exploring the AGN population at $z\lesssim 0.3$. Roughly half of all AGN to be found by \art\ are expected to be absorbed objects, including $\lesssim 10^3$ heavily obscured ones (those seen through intrinsic absorption columns $N_{\rm H}\gtrsim 10^{23}$~cm$^{-2}$). Many of these objects will not be detected in softer X-rays by \erosita, hence \art\ will provide a crucial contribution to the \srg\ census of AGN.

In preparation to the \srg\ mission, a series of simulations of \art\ scanning observations were done based on the preflight knowledge of the instrument's characteristics. It was demonstrated \citep{2018AstL...44...67M} that the sensitivity of \art\ observations should significantly depend on the intensity of charged particle background at the \srg\ orbit, whose preflight estimates varied by almost an order of magnitude. Accordingly, it was predicted that up to $\sim 10,000$ AGN might be detected in the course of the all-sky survey. As discussed below (\S\ref{s:bgr}), the actual in-flight background intensity proved to be relatively high (although lower than the most pessimistic expectations), which together with the preliminary counts of sources detected during the first two scans of the sky performed in December 2019--December 2020 (see \S\ref{s:allsky} below) suggests that $\sim 3,000$ AGN will be detected by \art\ after completion of its 4-year survey.

The resulting sample should significantly improve our understanding of AGN population properties. At the depth of $\sim 10^{-12}$~\flux, \art\ can find Seyfert galaxies (with typical X-ray luminosities $\lesssim 10^{44}$~erg~s$^{-1}$) out to $z\sim 0.3$, as compared to $z\lesssim 0.1$ accessible to current hard X-ray surveys by \textit{INTEGRAL} \citep{2020NewAR..9001545M} and \textit{Swift} \citep{2017ApJ...850...74K}. This will allow tracing the evolution of AGN (including obscured ones) over the last $\sim 3$ billion years.

The most interesting AGN found by \art\ will be studied in detail using follow-up observations in other wavebands and, possibly, pointed \srg\ observations after completion of the all-sky survey. Importantly, \erosita\ and \art\ together provide a broad spectral coverage from 0.2 to 30~keV for bright AGN. Another unique capability of the \srg\ mission is frequent and quasi-regular monitoring (during the all-sky survey) of faint X-ray sources located near the ecliptic poles. \art\ will probe AGN out to $z\sim 1$ in these regions (with a total area of a few hundred sq. deg), in the objects' rest-frame energy band $\sim 8$--24~keV.

\subsubsection{Clusters of galaxies}

It is expected that during the all-sky survey, \art\ will confidently detect in the 4--12~keV energy band $\sim 500$ most massive and hot low-redshift clusters of galaxies. For $\sim 100$ brightest ones, it will be possible to measure fluxes in the 6--12~keV energy band with $\approx 10\%$ precision. Combined with observations of the same clusters in the softer X-ray band, these hard-band measurements will significantly improve determinations of the intracluster gas temperatures. Moreover, \art\ should provide a few-$\sigma$ detections or tight upper limits on the hard X-ray flux for $\sim 1000$ clusters and unveil the possible presence of AGN in their central galaxies. This will lead to more reliable and tighter constraints on the cosmological parameters inferred from the \srg\ all-sky survey.

Deep pointed and scanning observations of selected galaxy clusters and groups with \art\ will enable mapping their hard X-ray emission at much larger angular scales than possible with \textit{NuSTAR}, and to search for non-thermal emission in clusters at these angular scales. This can be achieved thanks to the larger FOV ($36'$ in diameter) of \art\ compared to \textit{NuSTAR} ($13'\times13'$) and the more uniform background\footnote{\textit{NuSTAR} is known to suffer from stray light contamination (the light not focused by the optics), which produces a non-uniform background. This issue was properly taken into account during the \art\ design, which helped to significantly suppress stray light (\S\ref{sect:structure}). Additionally, \art\ scanning mode observations (\S\ref{sec:scans}) produce very uniform exposure coverage compared to tiling NuSTAR surveys.}.

\subsection{Galactic objects}

\subsubsection{X-ray binaries and cataclysmic variables}

Upon completion of its all-sky survey, \art\ is expected to uncover all sources with X-ray luminosity \Lx$\gtrsim 10^{34}$~\lum\ within 10~kpc from the Sun, i.e. throughout the Galactic bulge and over roughly half of the Galactic disk. This will represent a leap forward with respect to previous X-ray surveys, which have probed the same volume only down to \Lx$\sim 10^{35}$~\lum. There can be a sizeable Galactic population of low-mass and high-mass X-ray binaries with \Lx$\sim 10^{34}$--$10^{35}$~\lum. Despite a significant progress in the study of such systems in the last decades (e.g. \citealt{2008A&A...491..209R,2013MNRAS.431..327L,2016ApJ...825..132H,2019NewAR..8601546K,2020NewAR..8801536S,2020NewAR..9101544P}), their physical and statistical properties are still known much worse than those of brighter X-ray binaries with \Lx$\sim 10^{35}$--$10^{38}$~\lum. \art\ will make a census of such low-luminosity X-ray binaries for the first time. Using this unique sample, we hope to significantly improve our understanding of various regimes of accretion onto neutron stars and black holes. In addition, we expect to find a number of new significantly absorbed Galactic sources similar to those discovered with \int\ \citep{2015A&ARv..23....2W}.

The \art\ all-sky survey is also expected to provide a major contribution to the study of cataclysmic variables. Such objects (accreting white dwarfs) are believed to be much more numerous than neutron-star and black-hole X-ray binaries but their statistical studies have been severely limited by the insufficient sensitivity of previous X-ray surveys. We expect that up to $\sim 10^3$ CVs will be found by \art\ compared to just several dozens present in existing X-ray-selected samples (e.g. \citealt{2004A&A...423..469S,2013MNRAS.432..570P,2020AdSpR..66.1209D,2020NewAR..9101547L}). This together with the availability of accurate distances for many of these objects from the \textit{Gaia} mission \citep{2018A&A...616A...1G} will enable a precise measurement of the X-ray luminosity function of CVs for the first time \citep{2019AstL...45...62M}.

\subsubsection{Supernova remnants}

SNRs are sources of thermal X-ray emission originating in shock-heated gas and non-thermal (synchrotron) emission caused by high-energy electrons moving in magnetic fields. Non-thermal X-ray emission provides important information about particle acceleration properties, magnetic field strengths and turbulence near SNR shock fronts. \art\ observations will enable high-quality morphology studies of Galactic SNRs to better constrain their thermal and non-thermal emission components.

\subsection{Galactic X-ray background}

Due to regular, multiply repeated scanning of the whole sky and stable particle background conditions at the L2 halo orbit, \art\ will obtain a unique, high-quality map of the sky in the 4--12~keV energy band. Not only will this map be useful for detection of individual sources of various types, as discussed above, but it should also reveal the large-scale hard X-ray emission of the Galaxy in unprecedented detail. This so-called Galactic Ridge X-ray Emission (GRXE) is mostly the superposition of numerous unresolved faint X-ray sources such as CVs and stars with active coronae \citep{2006A&A...452..169R,2009Natur.458.1142R}. There remain a number of open questions pertaining to the GRXE origin. In particular, we still poorly understand its composition and how it varies across the Galaxy (e.g. \citealt{2018PASJ...70R...1K,2019ApJ...884..153P}). Another important issue is whether truly diffuse X-ray emission, associated with the hot phase of the interstellar medium, contributes significantly to the GRXE.

The \art\ map will significantly improve on the currently best map of the GRXE obtained by the \textit{RXTE} observatory \citep{2006A&A...452..169R}. The latter has poor angular resolution ($\sim 1$~deg), which greatly complicates the separation of the GRXE and individual moderately bright ($\sim 10^{-11}$~\flux) X-ray sources. Thanks to \art's excellent angular resolution of $\lesssim 1$~arcmin, it will be possible to single out individual sources brighter than $\sim 10^{-12}$~\flux\ and to obtain a much cleaner and sharper map of the GRXE.

\subsection{Transients}

\art\ is perfectly suited for discovering and monitoring transient and strongly variable sources of various types, from short (seconds to thousands of seconds) X-ray bursts and flares and gamma-ray bursts (GRBs) to long lasted (hours to years) Galactic X-ray transients and strongly variable AGNs.

To monitor bright transient sources, we can use the fact that singly reflected events may fall onto the detector with offset angles up to $\mysim50'$. For this type of events the \art\ field of view is $\mysim2$~deg$^2$. Of course, true images of sources cannot be obtained there, but it is possible to use \art\ as ``concentrator'' and measure X-ray fluxes without imaging. In the survey mode, the \srg\ spacecraft is rotating with a period of 4 hours around the Z axis, which is pointed towards the Sun and moving approximately 1 deg per day following the Sun. Any celestial source crosses the $\diameter1.6$~deg \art\ field of view approximately 10 times. Thus, in the ``concentrator'' mode, bright transient sources can be monitored for at least 28--32~hours (see details in \citealt{2019ExA....48..233P}).

Although the probability to catch GRBs in \art's field of view is not high, bright GRBs can penetrate through the shielding material of the telescope and induce a signal on the detectors. Therefore, \art\ is expected to detect several GRBs per year and provide precise timing information for their localization by means of triangulation with other space observatories.

\subsection{Search for dark matter signal}

The high efficiency of \art\ for conducting X-ray surveys, in particular its ability to cover the whole sky, opens up a new opportunity for searching for X-ray lines from sterile neutrino, a candidate constituent of dark matter. The prospects of the \srg\ mission in searches for keV-scale mass sterile neutrino dark matter radiatively decaying into active neutrinos and photons have been recently investigated by \cite{2020arXiv200707969B}. It was demonstrated that \art\ data acquired in the energy range from 5 to 20 keV could provide more stringent constraints compared to those obtained with \nustar\  \citep{2016PhRvD..94l3504N,2017PhRvD..95l3002P,2019PhRvD..99h3005N,2020PhRvD.101j3011R}. Therefore, the \art\ telescope has a high potential in testing the sterile neutrino dark matter hypothesis.

\section{\art\ in-flight performance}
\label{s:inflight}

\subsection{First light}
\label{sec:fl}

On 30 July 2019, just 17 days after the launch, \art\ was pointed toward its first target in the sky, the famous X-ray binary Cen\,X-3, consisting of a rapidly rotating neutron star and a massive normal star. The images obtained with all seven \art\ modules in the 5--16~keV energy band (Fig.\,\ref{fig:art_int}) confirmed the high sensitivity and imaging capabilities of the telescope.

These observations were also used to check the timing capabilities of \art\ and the accuracy of the on-board time scale. For these purposes, simultaneous observations of Cen\,X-3 with the {\it INTEGRAL} observatory \citep{2003A&A...411L...1W} were organized. The data analysis showed the excellent operation of all \art\ modules, which clearly detected the pulsed emission with a period of $\simeq4.8$~s. The measured pulse period and profile in the 5--16 keV energy band proved to be in good agreement with those measured by the JEM-X telescope aboard {\it INTEGRAL} in the 5--20 keV energy band.

\subsection{Calibration and performance verification phase}

\begin{figure}
    \centerline{\includegraphics[width=0.3\textwidth]{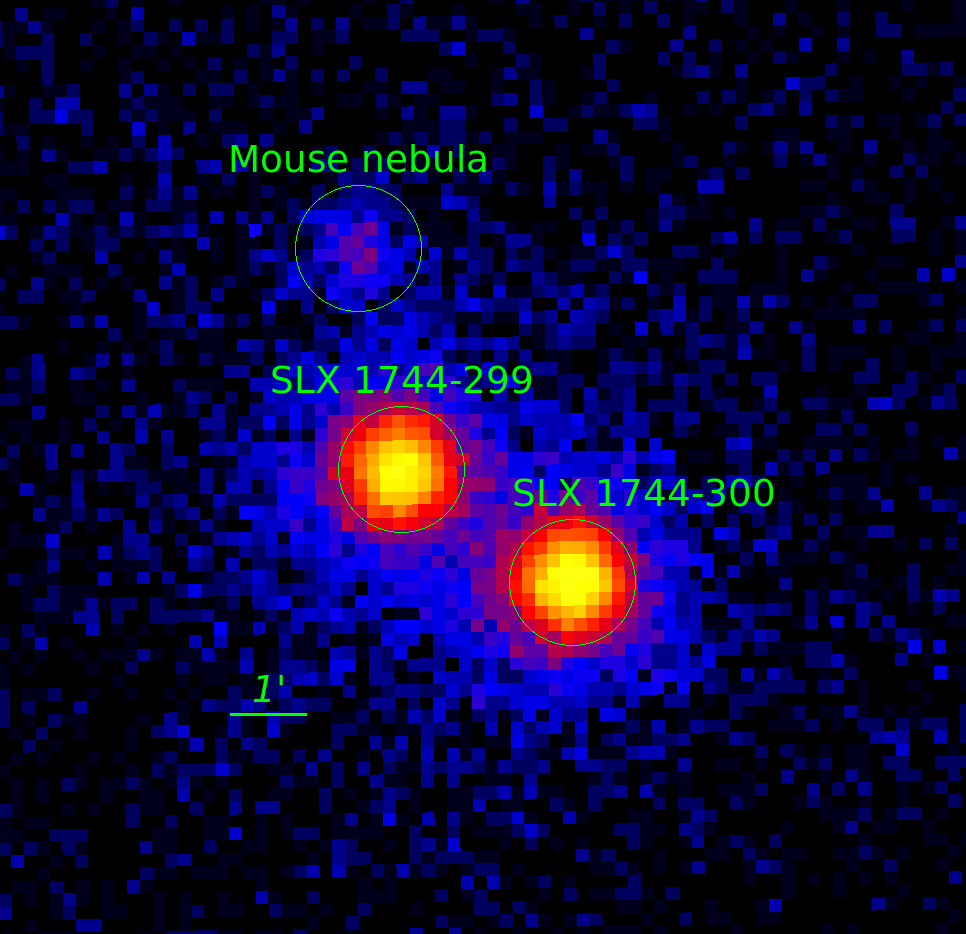}}
    \caption{Image of X-ray bursters SLX\,1744-299 and SLX\,1744-300 and the Mouse pulsar wind nebula, obtained with the \art\ telescope during the Galactic Bulge Survey (CalPV phase) in the 4--12~keV energy band. An angular distance between bursters are $\sim2.7$\arcmin.}
    \label{fig:psf_slx}
\end{figure}
\begin{figure}
    \centerline{\includegraphics[width=0.35\textwidth,viewport=10 166 564 685]{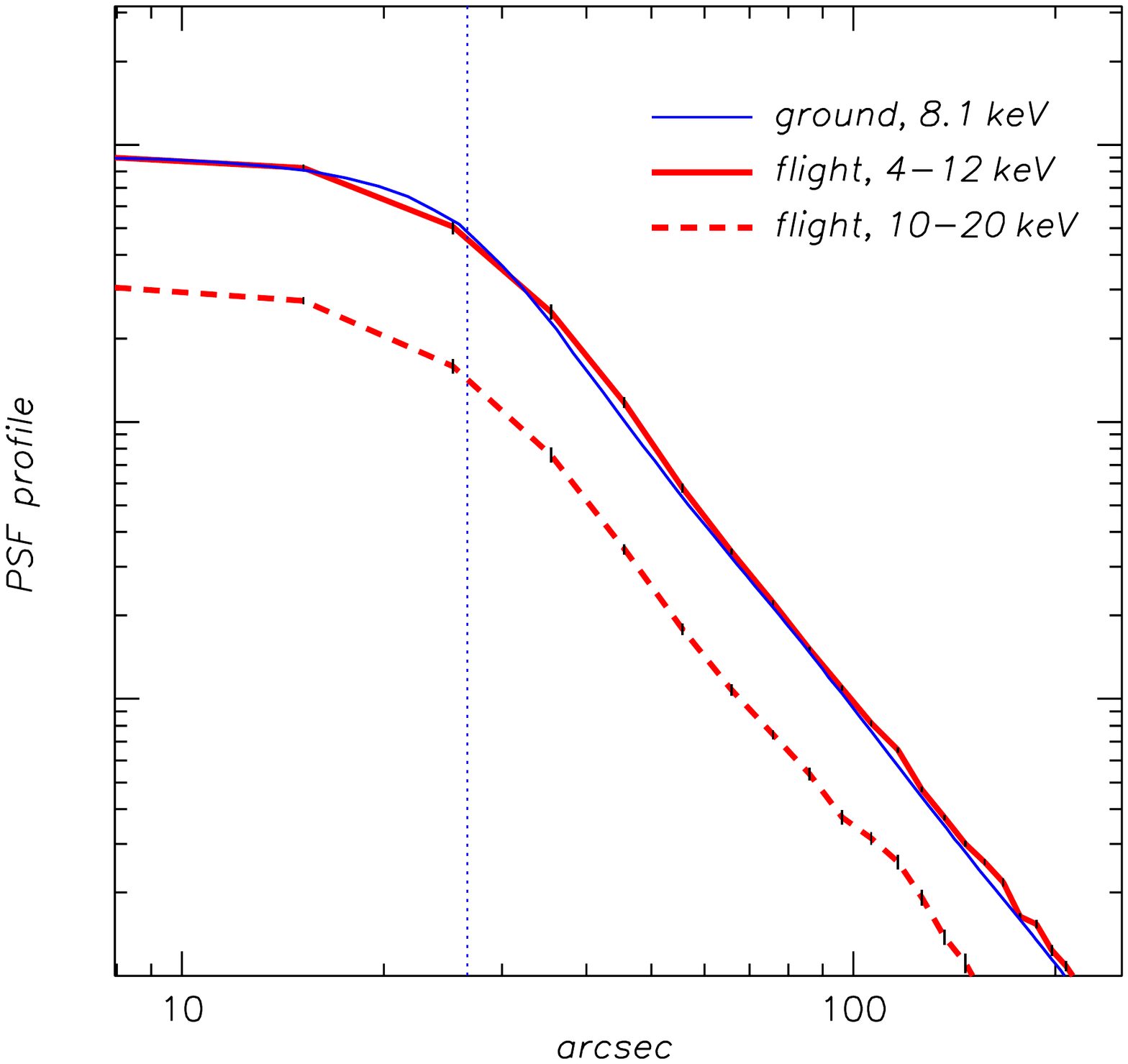}}
    \caption{PSF radial profiles averaged over \art's FOV in scanning mode: red solid and dashed lines --- observed for a bright X-ray source during the CalPV phase in the 4--12~keV and 10--20~keV bands, respectively, blue line --- obtained from ground calibrations (at 8.1~keV, see above). Uncertainties are shown by thin black lines. }
    \label{fig:psf_profile}
\end{figure}

The tuning and commissioning of all seven \art\ X-ray cameras (detectors) was completed on Aug. 25, 2019, after which a program of the \art\ calibration and performance verification observations began. This CalPV phase lasted for approximately one and half months and was followed by the CalPV phase of \erosita, during which \art\ was also operating.

During the CalPV phase, a number of bright X-ray sources were observed with \art\ in order to calibrate telescopes PSF, vignetting, effective area, timing capabilities, etc. All in-flight \art\ characteristics measured in these observations have proved to be close to those expected before launch (see details below). Moreover, \art\ revealed the transient X-ray pulsar GRO\,J1008-57 in the state with the lowest observed luminosity and significantly detected pulsations during this state for the first time \citep{2021arXiv210305728L}.

Additionally, a substantial part of the observational program during the CalPV phase consisted of surveys of specific regions of the sky (Galactic and extragalactic). By design, most of these surveys are much deeper than the \art\ 4-year all-sky survey is expected to be and thus significantly enhance its scientific value.

\subsubsection{Optical axes alignment and pointing accuracy}

It was found that, as expected, the boresights of individual MSs differ from each other by values of order of an arcminute or less. We have developed calibration procedures to measure the appropriate correction matrices, which are used for the alignment of the data obtained from different modules during ground data processing. For the initial calculation of boresight correction matrices, about 20 different pointed observations of bright X-ray sources from the CalPV phase were used. The accuracy of the current version of boresight corrections is about $6^{\prime\prime}$. This means that, given a star tracker data, we always know with 6 arcsec accuracy the position of a source in the field of view in any of the seven \art\ detectors and we can project an \art\ detector source image on the sky with this accuracy. The boresight corrections will be monitored and improved in the future.

Further analysis of CalPV data showed that the centers of the vignetting functions of the seven MSs do not coincide with the centers of the corresponding detectors. This implies that there is some misalignment between the optical axes of the MSs and that of the telescope. To measure the actual position of the telescope's optical axis at the detectors, a series of pointed observations of the Crab nebula was used. This showed that the optical axis position averaged over the seven MSs is about 2 arcmin away from the detector's center. Observations in the pointed mode are usually planned in such a way that the observed target is projected onto this position rather than the detector's center.

\subsubsection{Point spread function}

The observations of bright X-ray sources during the CalPV phase demonstrated that the in-flight point spread function (PSF) is very close to the expected one. To illustrate this, we show in Fig.~\ref{fig:psf_slx} a field containing X-ray bursters SLX\,1744-299 and SLX\,1744-300, located at a separation of $\sim2.7$\arcmin, and the Mouse pulsar wind nebula, as observed by \art\ during its ``Galactic Bulge Survey'' in the 4--12\,keV energy band. These observations were done in scanning mode, hence the PSF in this image is effectively averaged over the whole field of view (FoV) of \art. Similar PSF averaging takes place during the \art\ all-sky survey.

In Fig.~\ref{fig:psf_profile}, the PSF radial profile measured in scanning mode observations of a bright X-ray source (GX\,3+1) near the Galactic Center is compared to the radial profile of the PSF measured during the calibrations of the MSs at MSFC's Stray Light Test Facility \citep{2017ExA....44..147K}. To this end, the PSF measured at MSFC was convolved with a sliding box of the detector pixel size ($45^{\prime\prime}$) and then averaged over the FoV taking in account the vignetting function inferred from ray-tracing simulations. In both cases, the part of the source's flux at radii larger than $5^{\prime}$ was included in the background, which mimics the real situation corresponding to the detection of faint point sources.

The PSF profile measured in-flight during the CalPV phase and the one based on ground calibrations and simulations agree with each other within 10\%, as demonstrated in Fig.~\ref{fig:psf_profile}. The \art\ PSF FWHM for scanning and survey observing modes evaluated from the data of ground calibrations and simulations is $\approx 53^{\prime\prime}$. The actually measured in-flight PSF FWHM proves to be nearly the same.

\subsubsection{Effective area}

\art's effective area and vignetting function were measured using ground calibrations of mirrors and detector units and extensive ray-tracing simulations (see \S\ref{s:groundcal} above). In order to verify these measurements, a series of observations of the Crab nebula was carried out during the CalPV phase. The effective area was inferred to be very close to that obtained in simulations and ground calibrations. The analysis of these in-flight calibration data is continuing.

\subsubsection{Spectral resolution and energy scale}

\begin{figure}
\centerline{
\includegraphics[width=0.98\columnwidth,clip]{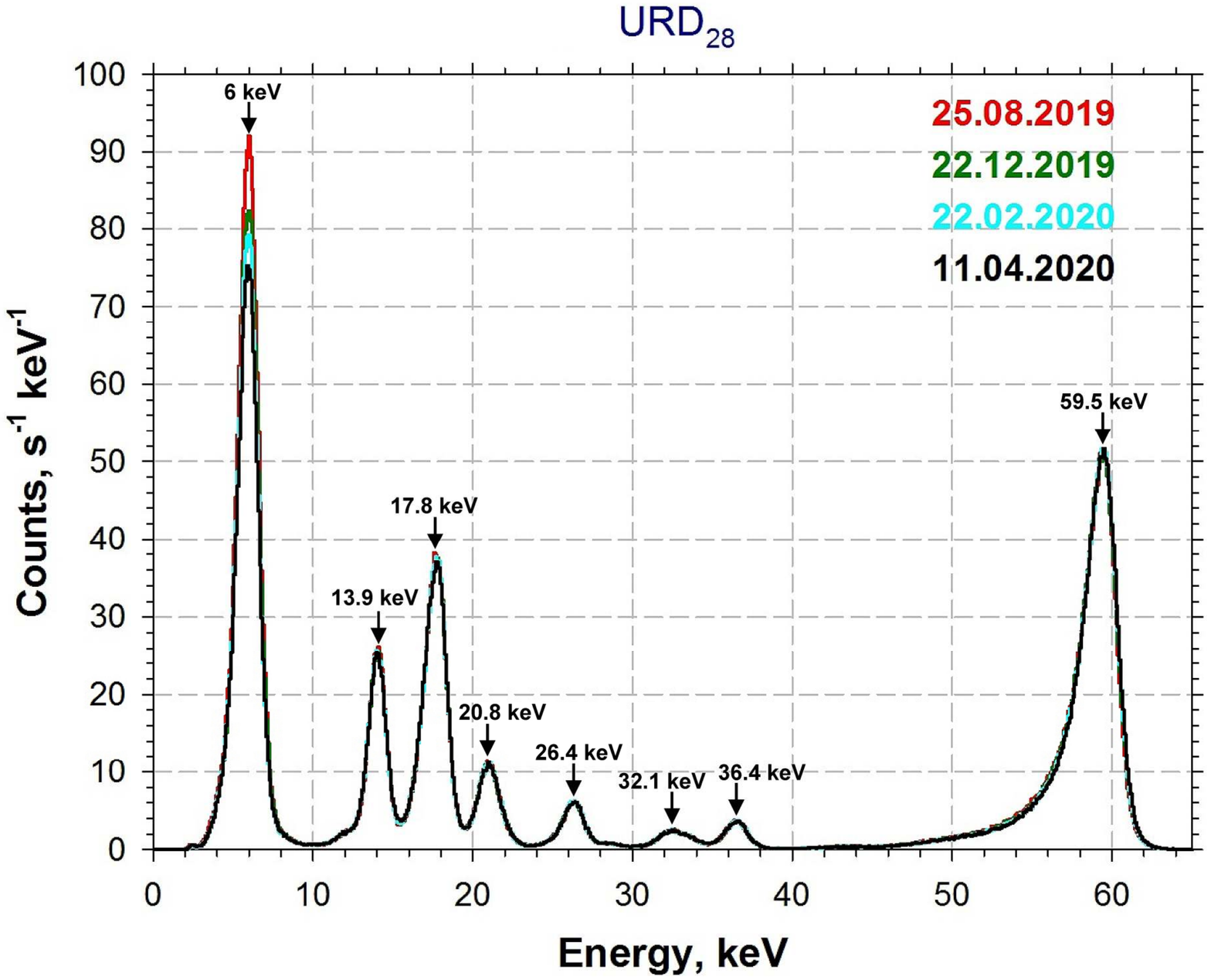}
}
\caption{Calibration spectrum of the $^{241}$Am+$^{55}$Fe source. The main calibration lines are indicated. The observed decrease of the intensity of the 5.9 keV line is related to the short half-life of Fe$^{55}$.}
\label{fig:cal_spc}
\end{figure}

In order to monitor the health of the detectors and to calibrate their energy scale and spectral resolution during the flight, \art\ uses its internal calibration source (see \S\ref{subsec:calsource} for details). Calibrations are performed regularly, every few months.

Figure~\ref{fig:cal_spc} shows the energy spectra obtained with one of the detectors in four such calibrations, performed on 25 Aug. 2019, 22 Dec. 2019, 22 Feb. 2020 and 11 Apr. 2020. Both the energy resolution and detector efficiency are in good agreement with the pre-flight measurements and do not change significantly with time and accumulated dose. The observed decrease of the intensity of the 5.9 keV line is related to the short half-life of Fe$^{55}$ (2.737~years).

\begin{figure}
\centerline{
\includegraphics[width=0.98\columnwidth,clip]{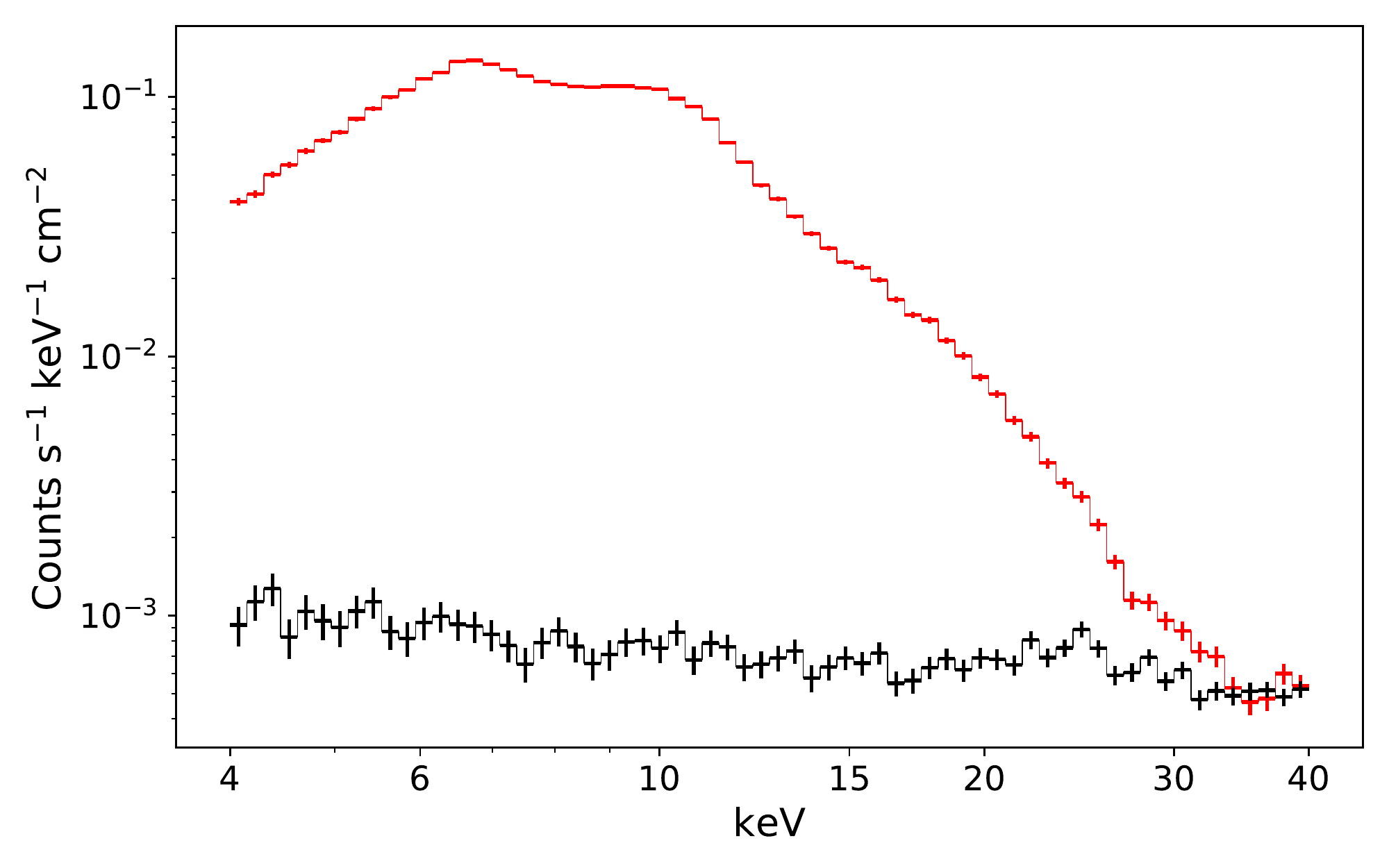}
}
\caption{Raw detector spectrum from a 20~ks observation of the bright Galactic HMXB OAO\,1657$-$415. }
\label{fig:oao_spe}
\end{figure}

As has already been noted, the \art\ MSs provide non-zero effective area at energies as high as $\approx35$~keV. To illustrate this, in Fig.~\ref{fig:oao_spe} we present the sum of 7 detector spectra of the bright Galactic HMXB OAO\,1657$-$415 accumulated over a 20~ks observation, along with the background extracted from an empty field. It is clearly seen that \art\ detects the source up to 30~keV. While the overall spectral shape is determined by the MS response, the broad peak at $\approx6.5$~keV reveals the presence of a Fe K$_{\alpha}$ line in the source's spectrum.

\subsubsection{Timing capabilities}

\begin{figure}
\centerline{
\includegraphics[width=0.98\columnwidth,clip]{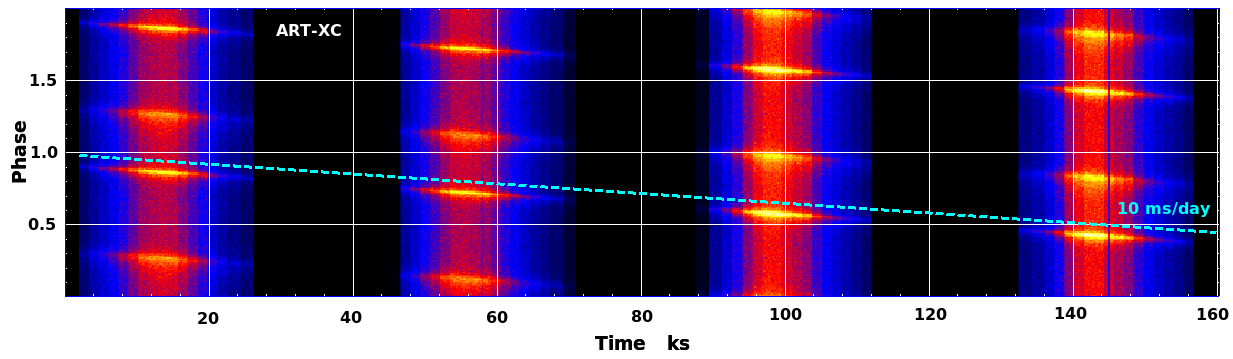}
}
\caption{Crab pulsar's dynamically folded lightcurve. The observed phase drift is caused by a deviation of the on-board clock frequency from the nominal value.}
\label{fig:crab_din_fold}
\end{figure}

\art\ has sub-millisecond time resolution. To test the timing capabilities of the telescope, we exploited a series of Crab pulsar observations. Using the Jodrell Bank's radio ephemeries we clearly detected $\sim 33$~ms pulsations from the source up to 30~keV. A further phase-connected analysis revealed that the frequency of the on-board clock is slightly lower than the nominal value, which leads to a lagging of the on-board clock relative to the Universal time at approximately 10~ms per day (see Fig.~\ref{fig:crab_din_fold}). This has also been confirmed based on measurements from the Ground Based Control Stations. By taking this systematic effect into account, \art\ can be used to perform a phase-resolved spectroscopy of pulsars with periods down to a few milliseconds.

One of the unique features of the \art\ telescope is the capability of its CdTe detectors and on-board electronics to process high photon fluxes. For example, during scanning observations of the very bright X-ray source Scorpion X-1 in September 2019  the \art\ detectors registered a count rate $\sim500$~cts\,s$^{-1}$ without any significant pileup. However, the estimated loss of the count rate due to the \art\ detector's dead time (0.77~ms) was about 37\% in this observation.

The count rate transferred from each detector in telemetry is 10--15~cts\,s$^{-1}$. If there are no bright X-ray sources in the telescope's field of view, the average count rate in the 4--100~keV energy band is $5.5\pm0.5$~cts\,s$^{-1}$ per detector.

\subsection{Detector background at the L2 halo orbit}
\label{s:bgr}

\srg\ is the first X-ray observatory near the L2 point, and there are no previous data on the background conditions in this location. Two main unknown factors could strongly affect the performance of X-ray telescopes such as \art: (i) the mean background level, which limits the all-sky survey sensitivity, and (ii) background stability on short timescales, i.e. the presence of flares, such as observed by many X-ray missions. Flares could spoil pointed observations and add a significant amount of noise for parts of the all-sky survey, making it less uniform.

\begin{figure}
\centerline{
\includegraphics[width=\columnwidth]{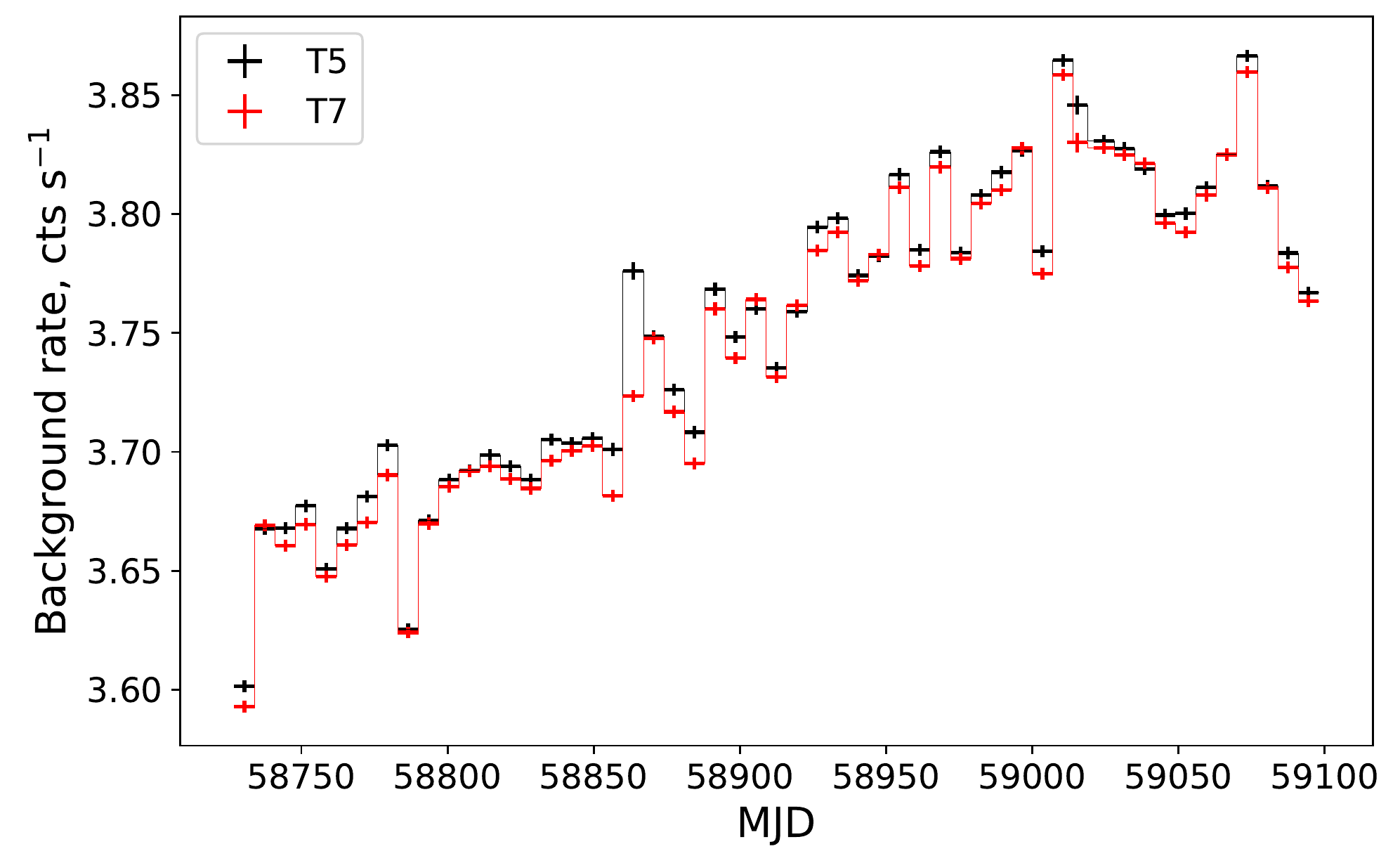}
}
\caption{Weekly-averaged rate of events in the 40--100 keV~band, which is dominated by charged particles. Two telescope modules are shown.}
\label{fig:art_bkg}
\end{figure}

The detectors of \art\ work up to energies of $\sim 100$~keV, while the mirrors have almost zero effective area above 35~keV. This makes it possible to use the count rate above $\sim 40$~keV as a proxy for the flux of cosmic rays. Figure~\ref{fig:art_bkg} shows the weakly-averaged count rate in the 40--100~keV band during the period from September 2019 to September 2020 for two telescope modules T5 and T7. The background is extremely stable with week-to-week variations of just a few percent. There is also an indication of an upward trend, but more data are needed to characterize it in detail.

A preliminary analysis of all available data has revealed a lack of bright flares on short timescales. However, there were no major Solar events during the first year of the \srg\ mission. With the onset of Solar cycle 25 in the next years the situation might change dramatically.

\begin{figure*}
\centerline{\includegraphics[width=0.98\textwidth]{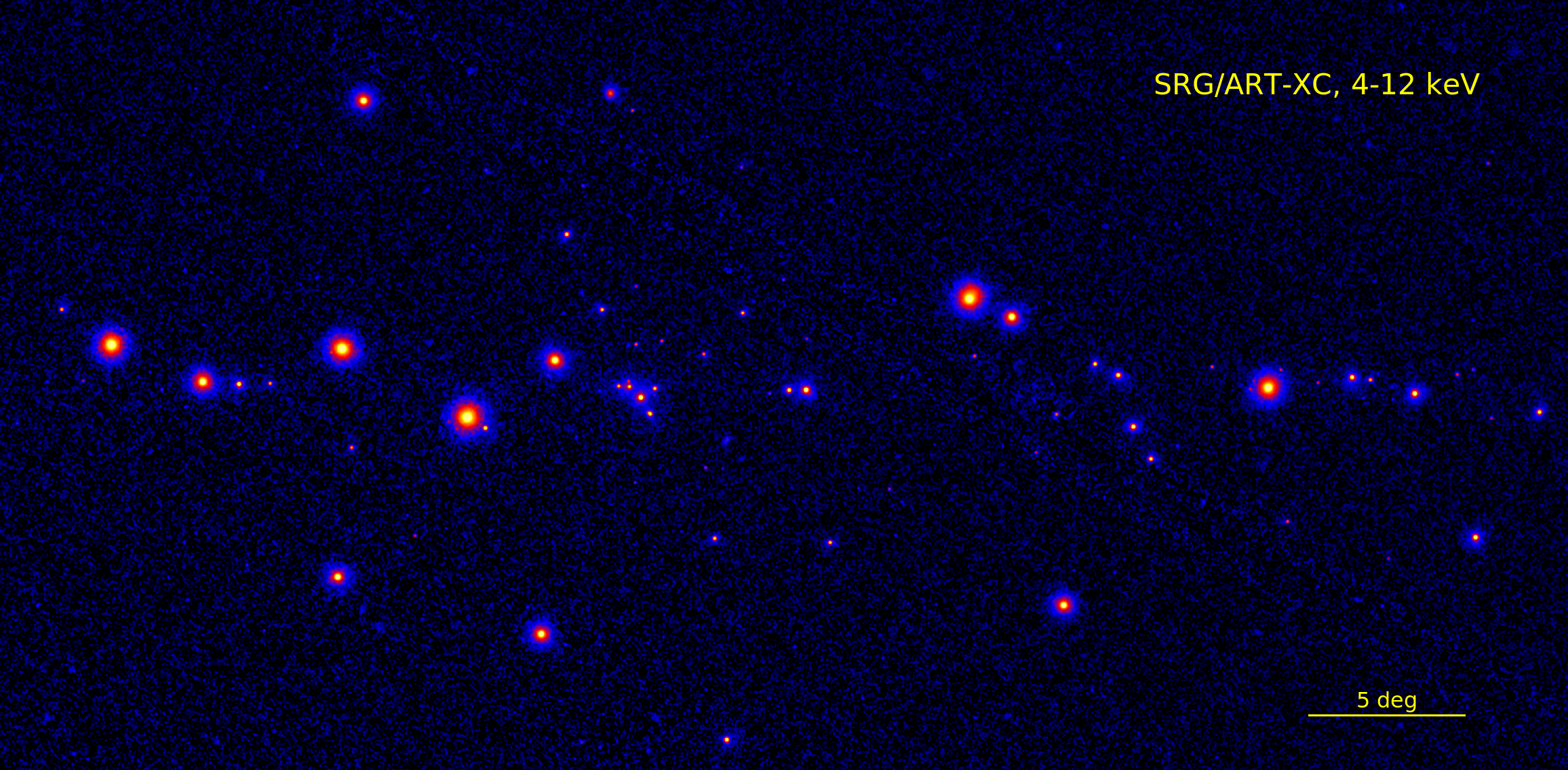}}
\caption{Map of the Galactic Plane region obtained during the first year of the \srg/\art\ all-sky survey in the 4--12~keV energy band. The region shown is of $\approx 50^{\circ}\times25^{\circ}$ size, centered on Galactic $l=355^{\circ}, b=0^{\circ}$, Aitoff projection. Vignetting corrected exposure vary approximately from 30\,s to 60\,s in different part of the image due to the specific survey strategy.}
\label{fig:art_allsky_survey}
\end{figure*}

\subsection{All-sky survey}
\label{s:allsky}

Upon completion of the CalPV phase, on 12 December 2019 the \srg\ observatory started its 4-year all-sky X-ray survey. By 10 June 2020 the entire sky had been covered by \art\ for the first time and by 15 Dec. 2020 for the second time.

Our preliminary analysis of the all-sky maps constructed from the \art\ data of the first year of the survey has revealed $\sim 700$ sources detected in the 4--12~keV energy band and $\sim 400$ sources detected in the 7--12~keV energy band (of which $\sim 100$ sources are not detected in the softer 4--7~keV band). Note that at energies above 6--7~keV \art\ is more sensitive (in survey mode) than  \erosita\ (Sunyaev et al., in preparation). Based on our current knowledge of the instrument and in-flight background conditions, we can predict that after completion of the 4-year all-sky survey \art\ will detect $\sim 5000$ sources in the 4--12~keV band.

Figure~\ref{fig:art_allsky_survey} shows a $\sim 1000$~sq. deg fragment of the all-sky map along the Galactic Plane obtained by \art\ in the 4--12~keV energy band during its first scan of the sky. It nicely demonstrates the unique characteristics (hard X-ray band, all-sky coverage, good angular resolution, high dynamic range, and uniformity) of the on-going \art\ survey.

\section{Summary}

The \art\ telescope on board the \srg\ observatory has been operating in orbit for more than a year and a half as at the time of writing, and its performance fully meets expectations. All the systems of the telescope are healthy and operating nominally. The series of observations performed during the early, CalPV phase of the mission revealed that the in-flight characteristics of the \art\ mirror systems and detectors are very close to the expectations based on the results of ground calibrations.

The all-sky survey that started on 12 Dec. 2019 is continuing smoothly without interruptions, with scientific data being dumped to the \art\ Science Data Center at IKI in Moscow on a daily basis. The first two \art\ maps of the whole sky in the 4--12~keV energy band, based on the data of the first and the second 6-month surveys, as well as their sum, have already been obtained, and work is in progress on the first \art\ catalog of X-ray sources. \art\ has already discovered a few dozens of new X-ray sources (see, e.g., \citealt{2020ATel13415....1S,2021ATel14357....1S, mereminskiyat2019wey,2021AstL...47...89M}), and efforts are underway to identify them with the help of the optical telescopes belonging to the \srg\ mission ground segment.

The \srg\ all-sky survey is planned to be completed in Dec. 2023, and there are good reasons to be optimistic that \art\ will provide a uniquely deep and sharp map of the entire sky in the 4--12~keV energy band. That will not, however, be the end of the mission, as the current \srg\ mission plan foresees the subsequent few years to be devoted to an extensive program of pointed observations of the selected astrophysical objects. During this post-survey phase, the full energy band of \art\ (4--30~keV) will be fully exploited for the spectroscopy of X-ray sources. The data of the ART-XC all-sky survey will eventually be made public. The date of the data opening is currently under discussion.

\section*{Acknowledgements}
The {\it Mikhail Pavlinsky} \art\ telescope is  the  hard X-ray instrument on board the \srg\ observatory, a flagship astrophysical project of the Russian Federal Space Program realized by the Russian Space Agency, in the interests of the Russian Academy of Sciences. The \art\ team thanks the Russian Space Agency, Russian Academy of Sciences and State Corporation Rosatom for the support of the \srg\ project and \art\ telescope. We thank the Lavochkin Association (NPOL) with partners for the creation and operation of the \srg\ spacecraft (Navigator) with an especial gratitude to E. Filippova, A. Pogodin and P. Merkulov. We thank the Acrorad Co., Ltd. (Japan), which manufactured the CdTe dies and Integrated Detector Electronics AS -- IDEAS (Norway), which manufactured the ASICs for the X-ray detectors. We thank our colleagues who have provided a crucial contribution to the ART-XC telescope but to our great regret have passed away: Valery Akimov (IKI) -- X-ray detectors scientist, Oleg Kozlov (IKI) -- designer of X-ray detectors mechanical parts, and Mikhail Gubarev (MSFC) -- designer of X-ray mirrors.


\bibliographystyle{aa} 
\bibliography{artxc_review}

\end{document}